\documentclass[floatfix,aps,amsmath,prd,twocolumn,eqsecnum,showpacs,nofootinbib,reprint]{revtex4-1}
\usepackage{enumerate,graphicx,graphics,amsfonts,amsmath,amssymb,bm,hyperref,psfrag,natbib,dcolumn}

\allowdisplaybreaks

\newcommand{\be}{\begin{equation}}
\newcommand{\ee}{\end{equation}}

\newcommand{\bs}{\begin{subequations}}
\newcommand{\es}{\end{subequations}}

\begin{document}
\title{Conservative self-force correction to the innermost stable circular orbit: comparison with multiple post-Newtonian-based methods}
\author{Marc Favata}
\thanks{NASA Postdoctoral Fellow}
\email{favata@tapir.caltech.edu}
\affiliation{Jet Propulsion Laboratory, 4800 Oak Grove Drive, Pasadena, California 91109, USA}
\thanks{Copyright 2010 California Institute of Technology. Government sponsorship acknowledged.}
\affiliation{Theoretical Astrophysics, 350-17, California Institute of Technology, Pasadena, California 91125, USA}
\date{26 August 2010}
\begin{abstract}
Barack and Sago [Phys.~Rev.~Lett., {\bf 102}, 191101 (2009)] have recently computed the shift of the innermost stable circular orbit (ISCO) of the Schwarzschild spacetime due to the conservative self-force that arises from the finite-mass of an orbiting test-particle. This calculation of the ISCO shift is one of the first concrete results of the self-force program, and provides an exact (fully relativistic) point of comparison with approximate post-Newtonian (PN) computations of the ISCO. Here this exact ISCO shift is compared with nearly all known PN-based methods. These include both ``nonresummed'' and ``resummed'' approaches (the latter reproduce the test-particle limit by construction). The best agreement with the exact (Barack-Sago) result is found when the pseudo-4PN coefficient of the effective-one-body (EOB) metric is fit to numerical relativity simulations. However, if one considers uncalibrated methods based only on the currently known 3PN-order conservative dynamics, the best agreement is found from the gauge-invariant ISCO condition of Blanchet and Iyer [Classical Quantum Gravity {\bf 20}, 755 (2003)], which relies only on the (nonresummed) 3PN equations of motion. This method reproduces the exact test-particle limit without any resummation. A comparison of PN methods with the ISCO in the equal-mass case (computed via sequences of numerical relativity initial-data sets) is also performed. Here a (different) nonresummed method also performs very well (as was previously shown). These results suggest that the EOB approach---while exactly incorporating the conservative test-particle dynamics and having several other important advantages---does not (in the absence of calibration) incorporate conservative self-force effects more accurately than standard PN methods. I also consider how the conservative self-force ISCO shift, combined in some cases with numerical relativity computations of the ISCO, can be used to constrain our knowledge of (1) the EOB effective metric, (2) phenomenological inspiral-merger-ringdown templates, and (3) 4PN- and 5PN-order terms in the PN orbital energy. These constraints could help in constructing better gravitational-wave templates. Lastly, I suggest a new method to calibrate unknown PN terms in inspiral templates using numerical-relativity calculations.
\end{abstract}
\maketitle
\section{\label{sec:intro}Introduction and motivation}
The primary purpose of this study is to compare recent gravitational self-force (GSF) calculations of the innermost stable circular orbit (ISCO) \cite{barack-sago_isco,barack-sago-circselfforcePRD2007,barack-sago-eccentricselfforce} with nearly all post-Newtonian (PN) and effective-one-body (EOB) methods. The first half of this paper provides introductory material, reviews previous related work, and summarizes the various PN/EOB approaches. Readers wishing to skip this material can proceed directly to the results in Sec.~\ref{sec:results}.
\subsection{\label{sec:GRsolnsintro}Regimes of the relativistic two-body problem}
One of the goals of this study is to provide insight on the various methods used to solve the relativistic two-body problem for the purpose of generating gravitational-wave (GW) templates.
We begin by briefly reviewing these methods.

The post-Newtonian (PN) approximation iteratively solves Einstein's equations using the approximation that a binary's relative orbital speed $v$ is small compared to the speed of light $c$. The PN equations of motion are known completely\footnote{Partial results at higher orders include the 4PN tail contribution \cite{blanchet-PRD1993-time-asymmetric} and the 4.5PN radiation-reaction terms \cite{gopakumar-iyer-iyer-PRD1997,*gopakumar-iyer-iyer-PRD1997-erratum}.} to 3.5PN order [i.e., computed to order $(v^2/c^2)^{3.5}$ beyond the Newtonian terms; see \cite{blanchetLRR} for references and a review]. For a binary with masses $m_1\leq m_2$ and $v/c\ll 1$, the PN approach is valid for any mass ratio $q\equiv m_1/m_2 \leq 1$, although it is known to ``converge'' more slowly if $q\ll 1$ \cite{cutler-etal-last3minutesPRL1993,poisson-bhpertVI-PNaccuracy,*poisson-bhpertVI-PNaccuracy-erratum,simone-leonard-poisson-will-CQG1997,leonard-poisson-CQG1998,yunes-berti-PNaccuracy,blanchet-PNaccuracy-confproc}.

When the binary separation is small and $v/c\sim 1$, the PN approximation breaks down, and other methods must be applied. One such method is numerical relativity (NR), the numerical solution of Einstein's equation without approximation. This approach has had much recent success (see \cite{pretorius-BBHreview,hannam-CQG-NRDA08review,sperhake-bookchapter2009,hinder-CQG-NRDA09review} for reviews), but computational limitations currently restrict it to modeling binaries with mass ratios $q\gtrsim 0.1$ \cite{gonzalez-etal-PRD2009} (however, see Refs.~\cite{lousto-etal-NR-IMRI-PRL2010,lousto-etal-IMRBBH-2010,lousto-zlochower-EMRBBHNR} for recent progress). For smaller mass ratios the time to inspiral increases like $T_{\rm insp} \sim 1/q$, and multiple spatial scales ($\Delta r \sim m_2$ and $\sim m_1=q m_2$) must be resolved accurately. This requires a finer spatial grid, smaller step-sizes, and longer evolution times. It will therefore be very difficult for NR to simulate more than a few orbits for binaries with very small mass ratios ($q\lesssim 10^{-2}$).

Because they will execute many observable orbital cycles in the highly relativistic ($v\sim c$) regime, an accurate description of extreme ($q\lesssim 10^{-4}$) and intermediate ($10^{-4} \lesssim q \lesssim 10^{-2}$) mass ratio binaries are amenable to neither PN nor NR methods. But they are amenable to a third method---the gravitational self-force approach. This is based on computing how a point-particle with mass $m_1\ll m_2$ deviates from geodesic motion around a black hole (BH) with mass $m_2$. The force that causes this deviation (the GSF; see \cite{poissonselforcereview,lousto-selfforcereview,detweiler-selfforcelec,barack-selfforcereview} for reviews and references) arises from the particle's own gravitational field. The GSF is responsible for \emph{dissipative effects} like the radiation-reaction force that causes the point-particle to lose energy and angular momentum to GWs as it inspirals. It is also responsible for \emph{conservative effects} which are time-symmetric and preserve the orbit-averaged constants of the motion.

One example of a conservative GSF effect is the shift in the periastron advance angle per orbit $\Delta \varphi$ due to finite-mass ratio corrections: e.g., at leading PN order the periastron advance can be written as \cite{DGI}
\be
\label{eq:periastronPN}
\frac{\Delta \varphi}{2\pi} \equiv k = \frac{3 (m_2 n)^{2/3}}{(1-e_t^2)} \left[1 + \frac{2}{3}q + O(q^2) \right],
\ee
where the mean motion is $n\equiv 2\pi/P_{\rm orb}$, $P_{\rm orb}$ is the periastron-to-periastron period, and $e_t$ is the ``time'' eccentricity appearing in the quasi-Keplerian formalism of \cite{DGI}. The first term represents the geodesic contribution; the $O(q)$ term represents the first-order conservative GSF correction.
Conservative GSF calculations of the periastron shift are discussed in \cite{barack-damour-sago_periastron,barack-sago-periastron}.
Another example of a conservative GSF effect is the finite test-mass shift in the frequency of the ISCO (which is the focus of this study).

While evaluating the full GSF has proven to be technically difficult, in the past three years four independent groups \cite{barack-sago-circselfforcePRD2007,detweiler-circselfforcePRD2008,berndtson-selfforcePhDthesis,sago-barack-detweiler,keidl-etal-GSFkerrI,shah-etal-conservGSF} have succeeded in computing the GSF for circular geodesics in Schwarzschild.\footnote{The GSF research program is strongly motivated by the need to produce accurate waveforms for \emph{extreme-mass-ratio inspirals} (EMRIs), an important source for the LISA mission \cite{lisaweb} consisting of a stellar-mass compact object inspiraling into a massive BH \cite{amaro-etal-emri-review-CGQ2007}.} More recently, Barack and Sago (BS) have computed the GSF for eccentric (bound) geodesics in Schwarzschild \cite{barack-sago_isco,barack-sago-eccentricselfforce}. They were then able to compute the change in the ISCO radius and angular frequency due to the conservative-piece of the GSF. This represents a significant milestone in (typically gauge-dependent) GSF calculations since the ISCO shift is a well-defined and easily understood strong-field quantity that can be compared with other approaches. As previous studies \cite{buonanno-cook-pretorius,boyle-etal-PRD2007,mroue-kidder-saul-PRD2008,boyle-etal-Efluxcomparison,baker-etal-PRL2007-NRPN,baker-etal-PRD2007-NRPN,hinder-etal-eccentricPN-NR,campanelli-etal-spinning-NR-PN-compare,hannam-etal-PN-NR-meet,berti-etal-multipolarnonspinning,berti-etal-mulitpolarspinning,hannam-gopa-NRPN-spin,damour-nagar-jena,damour-nagar-AEI,damour-nagar-caltechcornell,damour-nagar-PRD09,gopa-jena-eccentricPNNR,pan-buonanno-baker-etal-NRPN,buonanno-pan-baker-etal-nonspinningEOB,buonanno-caltechEOB09,yi-etal-spinningEOBcaltech-PRD2010}  have investigated the agreement between NR and PN-based waveforms in the $q\sim 1$ regime, the objective of this study is to further compare PN-based approaches with these new GSF results (which are exact in the $q\ll 1$ limit).

While we investigate several PN-based approaches below, let us briefly highlight the effective-one-body \cite{EOB-BD1,EOB-BD2,EOB-damour-lecnotes,damour-nagar-EOBlecnotes2009} approach, which has especially motivated this study. The EOB formalism attempts to improve the convergence of the PN two-body equations of motion by mapping the PN two-body Hamiltonian for the masses $m_1$ and $m_2$ to an ``effective'' Hamiltonian that (at the 2PN level) describes a particle with reduced mass $\mu=m_1 m_2/M=\eta M$ moving on geodesics of a ``deformed'' Schwarzschild background associated with a mass $M=m_1+m_2$.\footnote{Recall that $\eta=m_1 m_2/M^2=q/(1+q)^2 \leq 1/4$ is the reduced mass ratio, and is denoted $\nu$ by some authors.} (At the 3PN level the effective Hamiltonian must be supplemented with additional terms that do not arise from the Hamiltonian of the effective metric \cite{damour-jaranowski-schafer-EOBisco}.) The effective Hamiltonian describes the full conservative dynamics of the ($\mu, M$) binary, while quantities like the ``$\eta$-deformed'' ISCO, light-ring, and horizon depend only on the time-time piece of the EOB effective metric function $g_{tt}^{\rm eff} = -A(r)$.  To include the effects of dissipation, the EOB approach incorporates information from the PN expansion of the energy flux (resummed by various means to improve convergence). Using these elements, the EOB formalism is able to describe not only the inspiral, but also the transition region where the inspiral ceases to be adiabatic and the point mass $\mu$ begins to ``plunge'' into the BH with mass $M$. By matching to a sum of quasinormal modes (whose complex frequencies are determined by the mass and spin of the BH merger remnant) near the ``$\eta$-deformed'' light-ring associated with the EOB metric, the EOB approach produces a complete waveform that describes the inspiral, merger, and ringdown phases of binary BH coalescence.

The EOB formalism is also highly modular: on top of the ``base''-EOB (consisting of the 3PN EOB effective Hamiltonian), various elements can be added and their parameters adjusted. For example, a ``pseudo-4PN'' term can be added to the EOB metric, and its coefficient can be adjusted to help improve agreement with NR simulations. One can also add correction terms or multiplicative factors to the dissipative dynamics or to the amplitude of the waveform modes. Some of these terms attempt to improve agreement with the exact NR results by making educated guesses about some of the (non-PN-expanded) physics implicit in the two-body dynamics; other terms contain additional free parameters that can be adjusted to agree with NR simulations. This modularity gives the EOB approach a great deal of power and allows it to match the time-domain NR waveforms with high accuracy \cite{buonanno-pan-baker-etal-nonspinningEOB,buonanno-caltechEOB09,yi-etal-spinningEOBcaltech-PRD2010,damour-nagar-AEI,damour-nagar-jena,damour-nagar-PRD09}. While the EOB formalism thus serves as a framework to generate fast waveforms that agree well with NR simulations, it is less clear if the EOB approach---via its mapping of the two-body PN dynamics onto motion in a deformed Schwarzschild background---provides a deeper understanding of the two-body dynamics than is already afforded by the ordinary PN equations of motion. This issue has been addressed in several papers by Blanchet \cite{blanchet-isco-PRD2002,blanchet-PNaccuracy-confproc,blanchetiyer3PN,blanchet-confproc2003b} and will be discussed in more detail below.

Recently, Yunes et al.~\cite{yunes-etal-EMRI-EOB} (see also \cite{yunes-GWnote}) have applied the EOB formalism to model EMRIs. They show that if three fitting parameters are introduced into the dissipative portion of their model, they can accurately match the waveforms generated by Hughes's BH perturbation theory code \cite{hughesI,*hughesI-erratum1,*hughesI-erratum2,*hughesI-erratum3,*hughesI-erratum4,hughesII} for quasicircular inspiral in the Schwarzschild spacetime (with errors of $\lesssim 0.1$ rad in the phase and $0.002$ in the fractional amplitude over a two-year integration). In retrospect, this agreement is not necessarily surprising because (i) in the $q\rightarrow 0$ limit, the conservative dynamics for the EOB and BH perturbation theory approaches are the same---circular geodesics in Schwarzschild; and (ii) the dissipative dynamics is solely described by the GW energy flux $dE/dt$, which is known analytically to 5.5PN order in the test-mass limit \cite{tanaka-tagoshi-sasaki-55PN-PThPh1996}. Combined with analytic BH absorption terms at 4PN order \cite{poisson-sasaki-bhpertV-BHabsorptionPRD1995,tagoshi-mano-takasugi-BHabsorptionPThPh1997} and adjustable parameters for the 6PN and 6.5PN terms in the flux, it is plausible to expect that such a high-order PN expansion of the energy flux (combined with Pad\'{e} resummation and the factorization of an adjustable pole parameter $v_{\rm pole}$) can be made to match well with the numerical BH perturbation theory computation of the energy flux. Nonetheless, the work in Ref.~\cite{yunes-etal-EMRI-EOB} is an important demonstration that just as the EOB approach has been shown to be adept at fitting the results of NR simulations, it can also be applied to fit the results of BH perturbation theory calculations (although it remains to be seen how well the approach will work for eccentric, inclined Kerr orbits, which are expected to be typical for EMRIs).

Yunes et al.~\cite{yunes-etal-EMRI-EOB,yunes-GWnote} also use their EMRI/EOB approach to investigate higher-order GSF contributions that are not contained in Hughes's BH perturbation code (which only accounts for the leading-order, orbit-averaged dissipative piece of the GSF). In particular they find a phase error (between the EOB and BH perturbation theory waveforms) of $\sim 3$ radians due to the conservative GSF terms in the EOB Hamiltonian, and $\lesssim 20$ radians when the higher-order dissipative GSF terms in the EOB dynamics are included (see Fig.~3 of \cite{yunes-GWnote}).  However, it is far from clear that the EOB approach can make accurate statements about higher-order GSF effects (as Yunes \cite{yunes-GWnote} himself notes), and part of the motivation of this work is to assess the degree to which the EOB formalism embodies conservative GSF effects.

By construction, the EOB formalism completely accounts for the conservative dynamics in the test-mass limit. The finite test-mass information it incorporates originates from the ordinary PN two-body dynamics, which is known to converge slowly in the small-mass-ratio limit. Therefore, even though the EOB conservative dynamics is exact in the $q\rightarrow 0$ limit, and matches NR calculations reasonably well in the equal-mass limit (with the help of extra ``flexibility'' parameters), it is not at all clear how accurate the conservative EOB dynamics is for small (but nonzero) values of the mass ratio $q$. This issue is investigated here in the context of the recent GSF ISCO calculations (see also related work by Damour \cite{damour-GSF}).

In particular, this study is especially interested in the performance of the ``base'' EOB 3PN Hamiltonian. As discussed here and in \cite{damour-GSF,barack-damour-sago_periastron}, additional parameters can be introduced in the EOB formalism and calibrated to reproduce the results of GSF calculations. However, these GSF calculations are themselves currently limited to computing first-order in $q$ corrections to geodesic motion. In some situations (e.g., intermediate-mass-ratio inspirals or IMRIs and possibly EMRIs) $O(q^2)$ corrections are expected to be important, and no ``exact'' numerical technique exists to treat this case (but see \cite{lousto-etal-NR-IMRI-PRL2010,lousto-etal-IMRBBH-2010,lousto-zlochower-EMRBBHNR} for a first attempt). In this case one cannot expect to be able to fully calibrate the EOB formalism. If one would like to apply the EOB approach to model IMRIs \cite{yunes-etal-EMRI-EOB}, it is important to gain insight into how the conservative EOB dynamics performs in the absence of any calibration to known numerical results.
\subsection{Previous self-force comparison studies}
Comparisons between PN results and BH perturbation theory calculations have a long history (see Refs.~16\mbox{--}23 of Ref.~\cite{blanchet-etal-selfforceI}), but these involve only dissipative self-force effects like the radiated energy and angular momentum. Conservative GSF effects have been computed only very recently, and the first comparisons with PN results were performed by Detweiler \cite{detweiler-circselfforcePRD2008}. In particular, Detweiler identified two well-defined, gauge-invariant quantities\footnote{The quantities are gauge invariant in the following sense: for quasicircular orbits in Schwarzschild, the quantities $u^{\varphi}_1$ and $u^t_1$ (and $\Omega\equiv u^{\varphi}_1/u^t_1$) are unchanged under an infinitesimal coordinate transformation $x^{\mu} \rightarrow x^{\mu} + \xi^{\mu}$ provided that (i) $k^{\alpha} \partial_{\alpha}\equiv \partial_t + \Omega \partial_{\varphi}$ remains a helical Killing vector in the perturbed spacetime on a dynamical (orbital) time, and (ii) that the gauge change preserves the reflection symmetry across the equatorial plane \cite{detweiler-circselfforcePRD2008}.} that could be calculated analytically in the PN approach and numerically in the GSF approach. These quantities are the angular frequency $\Omega$ of a particle on a circular orbit as measured by a distant observer, and the time-component of the particle's four-velocity $u^t_1\equiv dt/d\tau_1$. (The quantity $u^t_1$ can be identified with the redshift of a photon emitted by the particle and received by a distant observer on the $z$-axis perpendicular to the circular orbit.) While $\Omega$ and $u_1^t$ are themselves functions of gauge-dependent quantities (such as the orbital radius and the metric perturbation), one can calculate both quantities numerically for a particular choice of gauge in the GSF approach, and also analytically compute $u^t_1$ as a function of $\Omega$ in a PN analysis. The redshift function can be expressed as
\be
u^t_1 = \left[ -(g_{\alpha \beta})_1 v_1^{\alpha} v_1^{\beta} \right]^{-1/2},
\ee
where $v_1^{\alpha}\equiv dy_1^{\alpha}/dt = (1,v_1^i)$ is the PN coordinate velocity of the particle and $(g_{\alpha \beta})_1$ is the regularized metric evaluated at the particle's position.

Using the near-zone metric previously computed to 2PN order, Detweiler \cite{detweiler-circselfforcePRD2008} constructed the PN expansion of $u^t_1(y)$ [where $y\equiv (m_2 \Omega)^{2/3}$] and compared with his numerical GSF evaluation. He showed good agreement at the 2PN level, and made a prediction for the value of the 3PN coefficient in $u^t_1$ [Eq.~\ref{eq:ut-PN-expand} below]. In later work Blanchet et al.~\cite{blanchet-etal-selfforceI} extended the computation of  $(g_{\alpha \beta})_1$ to 3PN order and improved the accuracy of the GSF calculations reported in \cite{detweiler-circselfforcePRD2008}, finding excellent agreement between the 3PN coefficient and a fit of its value to the GSF numerical results. Their refined numerical GSF results motivated Blanchet et al.~\cite{blanchet-etal-selfforceII,blanchet-etal-selfforceIII-confproc} to further push their PN computations of $(g_{\alpha \beta})_1$ and $u^t_1(y)$ to higher orders. In particular they computed the logarithmic corrections to the near-zone metric at 4PN and 5PN orders (the nonlogarithmic corrections are more difficult to compute and remain unknown). Their result for $u^t_1$ (when expanded in powers of the mass ratio $q$) takes the form \cite{blanchet-etal-selfforceI,blanchet-etal-selfforceII}
\be
\label{eq:ut-q-expand}
u^t_1 = u_{\rm Schw}^t + q u_{\rm SF}^t + q^2 u_{\rm PSF}^t + O(q^3),
\ee
where the Schwarzschild result is known exactly,
\be
u_{\rm Schw}^t = (1-3y)^{-1/2},
\ee
and the leading-order self-force piece is given by
\begin{multline}
\label{eq:ut-PN-expand}
u^t_{\rm SF} = -y \left[ 1 + 2y + 5y^2 - \left(-\frac{121}{3} + \frac{41}{32}\pi^2 \right) y^3 \right.
\\
- \left(\alpha_4 - \frac{64}{5} \ln y \right) y^4 - \left(\alpha_5 + \frac{956}{105} \ln y \right) y^5
\\
- \left(\alpha_6 + \beta_6 \ln y \right) y^6 - \left(\alpha_7 + \beta_7 \ln y \right) y^7 + o(y^7) \bigg],
\end{multline}
where the unknown coefficients $\alpha_4$, $\alpha_5$, $\alpha_6$, and $\beta_6$ at 4PN, 5PN, and 6PN orders were determined by least-squares-fitting to the accurate GSF results (see Table V of \cite{blanchet-etal-selfforceII}; if $\beta_7$ is set to zero, a value for $\alpha_7$ was also determined, but including $\beta_7$ caused the fits to worsen). Their results show that the successive PN approximations smoothly converge to the exact GSF results (see Fig.~2 of Ref.~\cite{blanchet-etal-selfforceII}).
The post-self-force piece $u_{\rm PSF}^t$ was also calculated to 3PN order (and the logarithmic terms to 5PN order), but there is no second-order GSF formalism with which to compare them.

In addition to the comparisons of the GSF with the PN expansion for $u_1^t(y)$, Barack and Sago also computed $u_1^t(y)$ using their independent GSF code \cite{barack-sago-circselfforcePRD2007} and compared their results with Detweiler's code \cite{sago-barack-detweiler}. Although the two codes use different interpretations of the perturbed motion, different gauges for the metric perturbation, and different numerical techniques, their results for $u_1^t(y)$ agree to within numerical errors.

Recent work by Damour \cite{damour-GSF} investigated conservative GSF corrections in the EOB approach. Damour points out that GSF calculations can provide information on the two functions that appear in the EOB effective metric. These functions are expanded in Taylor series in $u=M/r$ (which are further resummed via Pad\'{e} approximants). While pseudo-4PN and 5PN terms in this series have been constrained by NR simulations \cite{pan-buonanno-baker-etal-NRPN,damour-nagar-PRD09}, Damour discusses how GSF calculations can similarly constrain higher-PN-order terms when these effective-metric functions are expanded in the small-$\eta$ limit. Some of these constraints arise from the conservative corrections to the ISCO computed by BS (this is discussed further in Sec.~\ref{sec:fit4pn} below). Damour \cite{damour-GSF} also investigates how orbits with small eccentricity, as well as a special class of zoom-whirl orbits, can further constrain parameters appearing in small-$\eta$ expansions of the EOB effective metric. The study presented here contains some overlap with Damour's work \cite{damour-GSF}, but here we focus primarily on comparisons of ISCO calculations with a large variety of PN-based methods in addition to the EOB approach.

Even more recently, Barack, Damour, and Sago \cite{barack-damour-sago_periastron} have computed the GSF correction to the rate of periastron advance in the small-eccentricity limit. They compare their numeric results with a particular gauge-invariant function $\rho(x)$, which is related to the ratio of the radial and azimuthal orbital frequencies and depends on the small-mass-ratio expansion of functions appearing in the EOB metric. They find very good agreement with the known 3PN expansion of $\rho(x)$, and are able to set constraints on higher-order nonlogarithmic terms in $\rho(x)$ (the 4PN and 5PN logarithmic terms having been recently computed in \cite{damour-EOBlogterms} and reported in \cite{barack-damour-sago_periastron}).
\subsection{\label{sec:isco-nr-compare}Previous comparisons of PN-based ISCO calculations with numerical relativity}
While comparisons between conservative GSF and PN results are very recent, comparisons between PN and NR results have a long history. Particularly relevant to this study are comparisons between PN-based ISCO calculations and earlier work in NR involving quasicircular initial data (QCID) calculations.
By numerically constructing sequences of quasicircular initial-data sets, several QCID studies computed the ISCO frequency for equal-mass binary BHs \cite{cook-effectivepotential-PRD1994,baumgarte-isco-PRD2000,pfeiffer-saul-cook-quascirc-PRD2000,gourgoulhon-etal-hkvII-PRD2002,baker-quasicircdata-grqc-2002,cook-pfeiffer-PRD2004,yo-cook-shapiro-baumgarte-PRD2004,tichy-brugmann-initialdata-PRD2004,hannam-quasicirc-PRD2005,alcubierre-etal-quasicirc-PRD2005,caudill-etal-initialdata-PRD2006}.
These calculations typically involve solving a subset of the full Einstein equations subject to certain approximations (such as the presence of a helical Killing symmetry, or specifying the conformal spatial metric to be flat or a linear superposition of two Kerr BHs.)
Several of these works also compared their results with PN ISCO estimates.\footnote{For even earlier work on the ISCO in comparable-mass binaries, see Refs.~\cite{clarkeardley,blackburn-detweiler-PRD1992}. Note also that Ref.~\cite{EOB-BD1} compared some PN and EOB ISCO methods, but did not include comparisons with numerical calculations.} The ISCO in the unequal-mass case has not been studied as thoroughly, but see Ref.~\cite{harald-thesis} for an extension of the work in \cite{pfeiffer-saul-cook-quascirc-PRD2000} to unequal-mass BH binaries.

Blanchet \cite{blanchet-isco-PRD2002} compared a variety of PN methods for computing the ISCO\footnote{Specifically, Blanchet \cite{blanchet-isco-PRD2002} considered the standard PN energy function, EOB, the $e$-method, and the $j$-method; these are discussed in detail below.} to the numerical result from \cite{gourgoulhon-etal-hkvII-PRD2002} for corotating, equal-mass binaries. However, aside from the PN energy-function approach (to which Blanchet \cite{blanchet-isco-PRD2002} derived the spin-corrections), the other ISCO estimates were computed only for nonspinning BHs, so it is unclear how to precisely evaluate all of the resulting comparisons. Damour et al.~\cite{damour-gourgoulhon-grandclement-ISCO-PRD2002} extended EOB computations of the ISCO to corotating binaries, and compared their calculations with QCID results from \cite{cook-effectivepotential-PRD1994,pfeiffer-saul-cook-quascirc-PRD2000,gourgoulhon-etal-hkvII-PRD2002} in the nonrotating case. In Table \ref{tab:iscoequalmass} below I give an update of these equal-mass comparisons, showing how a larger variety of PN methods compares with more recent QCID calculations for nonrotating BHs. My results are consistent with and expand on those presented in \cite{blanchet-isco-PRD2002,damour-gourgoulhon-grandclement-ISCO-PRD2002}.

In Refs.~\cite{blanchet-PNaccuracy-confproc,blanchet-confproc2003b} Blanchet reviews his results from \cite{blanchet-isco-PRD2002} and argues against the notion that the two-body problem in the comparable-mass limit is better represented by resummation methods (Pad\'{e} approximants and EOB) than by standard Taylor PN expansions.\footnote{Focusing on the energy flux rather than the ISCO, Mrou\'{e} et al.~\cite{mroue-kidder-saul-PRD2008} also examined the role of Pad\'{e} approximants versus Taylor expansions. They argue that Pad\'{e} approximants do not always help to accelerate the convergence of the energy flux in either the test-mass or equal-mass limits (although their Fig.~5 indicates that some Pad\'{e} approximants of the flux perform better than Taylor expansions). They also argue that the use of Pad\'{e} approximants in generating waveform templates does not offer significant advantages over using Taylor expansions.} His argument is based on an estimate of the radius of convergence of the PN expansion of the circular-orbit energy [Eq.~\eqref{eq:E3PN} below]. In the test-mass limit this radius of convergence is found to be  $x\equiv (M\Omega)^{2/3} = 1/3$, corresponding to the frequency of the Schwarzschild light-ring (the innermost location where circular orbits can exist). But in the equal-mass case this estimate of the convergence radius occurs at $x\approx 2.88$, implying that there is no notion of a deformed light-ring in the comparable-mass case. This is in contrast to the EOB approach, which describes the two-body problem as containing an $\eta$-deformed light-ring. Blanchet concludes that the PN two-body problem does not appear to be ``Schwarzschild-like.'' Blanchet also argues that the 3PN value of the ISCO frequency in the equal-mass ($\eta=1/4$) limit [$x_{\rm isco}(1/4) \approx 0.2$, as computed from the minimum of the orbital energy] is likely to be accurate because the ISCO lies well within the radius of convergence [$x(1/4)\approx 2.88$] of the PN series. This is in contrast to the test-mass limit, where the exact ISCO frequency $x_{\rm isco}^{\rm Schw} = 1/6$ is rather close to the light-ring $x_{\rm lr}^{\rm Schw}=1/3$.

In a more recent study of QCID, Caudill et al.~\cite{caudill-etal-initialdata-PRD2006} improve upon the previous work of \cite{cook-pfeiffer-PRD2004} and compare their QCID calculations to PN formulas for the ISCO. Their value for the ISCO frequency for nonspinning binaries [$M\Omega_{\rm isco}(1/4)\approx 0.12$] was found to agree more closely with a standard 3PN ISCO estimate \cite{blanchet-isco-PRD2002} [with $\approx 7\%$ error using the 3PN orbital energy; Eq.~\eqref{eq:E3PN} below] than with a 3PN EOB estimate \cite{damour-gourgoulhon-grandclement-ISCO-PRD2002} of the ISCO (with $\approx 36\%$ error; see Figs.~15\mbox{--}17 and Table II of \cite{caudill-etal-initialdata-PRD2006}, and Table III of \cite{hannam-quasicirc-PRD2005}). In the corotating case [$M\Omega_{\rm isco}^{\rm corot}(1/4)\approx 0.107$], the 3PN EOB ISCO \cite{damour-gourgoulhon-grandclement-ISCO-PRD2002} performed better ($\approx 9\%$ error versus $17\%$ for the standard 3PN case \cite{blanchet-isco-PRD2002}; see Table III of \cite{caudill-etal-initialdata-PRD2006}). These conclusions are qualitatively consistent with Fig.~3 of \cite{damour-gourgoulhon-grandclement-ISCO-PRD2002}. Comparisons between PN and QCID calculations of the equal-mass ISCO will be further addressed in Sec.~\ref{sec:results} (see Table \ref{tab:iscoequalmass} below).
\subsection{\label{sec:summary}Summary}
In Sec.~\ref{sec:iscointro} I briefly review the definition of the Schwarzschild ISCO and the difference between the ISCO and the ICO (innermost circular orbit), and give a short discussion of the GSF. I elaborate on how the dissipative self-force affects the ISCO, and then review the conservative ISCO shift calculations by BS. I also review Damour's \cite{damour-GSF} reformulation of the BS ISCO frequency shift into the standard PN notation and gauge.

Section \ref{sec:PNiscos} reviews all of the PN/EOB-based approaches for computing the ISCO: (1) Section \ref{sec:pnenergy} discusses the minimization of the standard 3PN energy function. Section \ref{sec:pneqns} examines (2) a stability analysis of the standard 3PN equations of motion, leading to two algebraic equations for the ISCO radius and frequency that must be solved numerically. It also discusses additional approaches from Blanchet and Iyer \cite{blanchetiyer3PN}. These involve expressing an analytic condition for the ISCO as a PN expansion in terms of either (3) the harmonic-gauge radial coordinate used in the Lagrangian form of the 3PN equations of motion or (4) the Arnowitt-Deser-Misner (ADM) radial coordinate used in the Hamiltonian formulation of the equations of motion. As discussed in \cite{blanchetiyer3PN}, these last two ISCO conditions can be reformulated in terms of (5) a single gauge-invariant analytic condition that can be solved for the ISCO frequency. This ISCO criterion [Eq.~\eqref{eq:C0invar}] is particularly special because (i) it contains the exact Schwarzschild ISCO without applying any resummation methods, and (ii) it produces the closest agreement to the BS conservative GSF ISCO shift of any 3PN-order method.
Section \ref{sec:pnhamiltonin} computes the ISCO by (6) numerically solving an algebraic system derived from the 3PN Hamilton's equations.

Section \ref{sec:hybridmethods} discusses a variety of ``hybrid'' methods that were originally inspired by Kidder, Will, and Wiseman (KWW) \cite{kidderwillwiseman-transition-PRD1993}. These hybrid methods all involve removing test-mass-limit terms in PN expressions and replacing them with the equivalent (fully relativistic) expressions from the Schwarzschild spacetime. In particular, this section considers (7) a hybrid 3PN energy function (Sec.~\ref{sec:hybridenergy}), (8) the KWW equations of motion, extended to 3PN order (Sec.~\ref{sec:hybridpneqns}), and (9) a Hamiltonian version of the KWW equations \cite{wex-schafer-CQG1993} (also extended to 3PN order; \ref{sec:hybridhamiltonian}). Section~\ref{sec:emethod} discusses two additional resummation approaches based on minimizing Pad\'{e} approximants of (10) an ``improved'' PN energy function (the \emph{$e$-method} \cite{damour-iyer-sathyaprakash-PRD1998}) and (11) a function based on the orbital angular momentum (the \emph{$j$-method} \cite{damour-jaranowski-schafer-EOBisco}).

Section \ref{sec:EOBmethods} discusses the EOB approach and, in particular, reviews the three ways of expressing the EOB effective-metric function $A(r)$ (which determines the EOB ISCO): (12) as a Taylor series expansion, (13) as a Pad\'{e} approximant of the Taylor series, and (14) via a new logarithmic expression introduced in \cite{barausse-buonanno-spinEOB}. Lastly, Sec.~\ref{sec:shanks} discusses (15) the Shanks transformation, a series acceleration method that is applied to several of the above PN-based ISCO calculations (which are themselves each computed at multiple PN orders).

Section \ref{sec:results} gives the results from these 15 ISCO calculations. These are provided in Table \ref{tab:compareisco}, which shows the $O(\eta)$ conservative correction to the exact Schwarzschild ISCO frequency, as well as its deviation from the exact BS result. Figure \ref{fig:dcOmega} illustrates a subset of this table in graphical form.  The ISCO in the equal-mass case is also tabulated, and compared with the QCID results from \cite{caudill-etal-initialdata-PRD2006}. Figure \ref{fig:iscofreq} shows the ISCO frequency for several of the considered methods as a function of $\eta$.

Section \ref{sec:discuss} draws a number of observations from Tables \ref{tab:compareisco} and \ref{tab:iscoequalmass}. \emph{Readers wishing to get to the main point of this paper as quickly as possible can skip directly to that section.} The most important points to take away are:
\begin{enumerate}[(a)]
\item The best agreement with the BS results comes from an EOB approach that includes a pseudo-4PN term that is calibrated to the comparable-mass Caltech/Cornell NR simulations in \cite{buonanno-caltechEOB09}.
\item If we do not consider calibrated methods but rely only on our current 3PN-level understanding of the conservative two-body problem, the best agreement with the BS ISCO shift comes from the gauge-invariant ISCO condition of Blanchet and Iyer \cite{blanchetiyer3PN}. This ISCO method is special because it already contains the exact Schwarzschild ISCO without introducing any ``manual'' resummation.
\item An extension of this gauge-invariant ISCO condition to spinning binaries shows that it also (i) reproduces the Kerr ISCO to the expected order in the spin parameter, and (ii) reproduces the conservative shift in the ISCO due to the spin of the test-particle. This is discussed further in a companion paper \cite{favata-PNspinisco}.
\item If we compare PN/EOB approaches with numerical relativity calculations of the ISCO based on  the quasicircular initial-data calculations of \cite{caudill-etal-initialdata-PRD2006}, then the ISCO computed from the standard 3PN energy function performs better than all EOB methods.
\item In both the extreme-mass-ratio and equal-mass cases, the 3PN EOB approach provides a single consistent method that computes conservative corrections to the ISCO in both the small-mass-ratio and equal-mass cases with nearly the same error ($\sim 27\%$). Calibrating a pseudo-4PN term to NR simulations or to the BS result reduces this error in both limits.
\end{enumerate}

Section \ref{sec:NR} discusses various ways that ISCO computations from GSF and NR can improve GW templates. Section~\ref{sec:fit4pn} discusses how these numerical ISCO calculations can be used to fit pseudo-4PN parameters appearing in the EOB effective metric. Section \ref{sec:phenomen} discusses how GSF results can be used to fix some of the free parameters that appear in phenomenological inspiral-merger-ringdown templates. Section \ref{sec:fitEpn} discusses how GSF and QCID ISCO results can constrain the undetermined functions appearing in the 4PN- and 5PN-order pieces of the PN orbital energy [Eq.~\eqref{eq:E3PN}]. Section \ref{sec:QCID} briefly introduces a new approach to combine results from full-NR evolutions with QCID calculations to help fix higher-PN-order terms in inspiral templates. Section \ref{sec:conc} discusses conclusions and future work.

Geometric units ($G=c=1$) are used throughout this work. Note that since formulas are used from a large number of references, similar or identical quantities are sometimes denoted differently in different subsections of this paper. I have largely tried to keep the notation consistent with the original source rather than unify the notation throughout the paper. The context hopefully makes the intended meaning clear.
\section{\label{sec:iscointro}The ISCO and its self-force corrections}
\subsection{\label{sec:iscoreview}The Schwarzschild ISCO}
The notion of an ISCO arises from the geodesic dynamics of a test particle in the Schwarzschild geometry (cf.~Ch.~25 of \cite{mtw}). In the Schwarzschild geodesic equations
\bs
\be
\label{eq:schwgeo-r}
\left(\frac{dr}{d\tau}\right)^2 = \tilde{E}^2 - V_{\rm eff}(r,\tilde{L}),
\ee
\be
\label{eq:schwgeo-phi}
\frac{d\varphi}{d\tau} = \frac{\tilde{L}}{r^2}, \;\;\;\;\;\;\;  \frac{dt}{d\tau} = \frac{\tilde{E}}{1-2m_2/r},
\ee
\es
the radial motion is governed by an effective potential
\be
V_{\rm eff} = (1-2m_2/r)(1+\tilde{L}^2/r^2),
\ee
where $\tilde{E}\equiv E/m_1$ and  $\tilde{L}\equiv L/m_1$  are the energy and angular momentum per unit test-mass, $\tau$ is proper time along the geodesic, and $(t,r,\varphi)$ are Schwarzschild coordinates. From the condition for circular orbits, $\partial V_{\rm eff}/\partial r=0$, one easily finds that the angular momentum $\tilde{L}_0$ and radius $r_0$ for circular orbits are
\be
\label{eq:Lschw}
\tilde{L}_0^2 = m_2 r_0^2/(r_0-3m_2) \;\;\; \text{and}
\ee
\be
r_0 = \frac{\tilde{L}_0^2}{2m_2} \left( 1 + \sqrt{1-12m_2^2/\tilde{L}_0^2} \right).
\ee
The latter expression indicates that there is a minimum angular momentum for circular orbits, $\tilde{L}_0^2 \geq \tilde{L}^2_{\rm crit} \equiv 12m_2^2$, and to this minimum there corresponds a radius $r_0=6m_2$ below which no \emph{stable} circular orbits can exist (an ISCO). [\emph{Unstable} circular orbits can exist below this radius (down to $r_0=3m_2$), but their angular momenta are greater than $\tilde{L}_{\rm crit}$.].

A critical radius for circular orbits can also be derived by finding the radius that minimizes the orbital energy of the test mass along a sequence of circular orbits. The energy along circular orbits is easily found by substituting $dr/d\tau=0$ and $\tilde{L}=\tilde{L}_0$ in Eq.~\eqref{eq:schwgeo-r}, yielding
\be
\label{eq:Eschw-1}
\tilde{E}_0^2 = \frac{(r_0-2m_2)^2}{r_0 (r_0-3m_2)},
\ee
from which one can easily verify that the energy minimum occurs at $r_0=6m_2$.

\subsection{\label{sec:ISCOvsISCO}ISCO vs ICO}
The critical point obtained by minimizing the energy of a circular orbit is sometimes referred to as the ICO (innermost circular orbit)\footnote{In Ref.~\cite{bcv1,*bcv1-erratum} the ICO is referred to as the MECO (maximum binding-energy circular orbit). The designation LSO (\emph{last stable orbit}) is used by some authors to refer to an ICO, and by others to refer to the generalization of the ISCO to generic orbits.} \cite{blanchet-isco-PRD2002}. To clarify, an ICO is \emph{defined} to be the frequency where the circular-orbit energy satisfies $dE/d\Omega =0$. The ISCO, on the other hand, refers to the point of onset of a dynamical instability in the equations of motion for circular orbits. In Schwarzschild the ICO and the ISCO are clearly the same. This is also true for Kerr and for the conservative orbital dynamics defined via the EOB approach. In Sec.~\mbox{IV~A~2} of Ref.~\cite{bcv1,*bcv1-erratum}, the authors show (in a Hamiltonian formulation) that the ISCO and ICO are formally equivalent if the Hamiltonian is known exactly. However, the ICO and ISCO yield different critical frequencies because the Hamiltonian (or, equivalently, the energy or the equations of motion) are known only to some finite (say $n$th) PN order. Hence the analytic conditions defining the ICO and ISCO, having different functional forms, differ (when PN-expanded) in terms at $n+1$ and higher PN orders. Since the conditions that define the ISCO or ICO are usually solved numerically, these implicit higher-PN terms (which would normally be truncated in an analytic expansion) cause the numeric values for the ISCO and ICO frequencies to differ. (See also Sec.~4.3 of \cite{schaefer-confproc-massmotion2009} for a related discussion.)

From a practical standpoint, both an ICO and ISCO signify a frequency at which the standard PN description of adiabatic circular orbits breaks down. In studies that consider the finite-mass-ratio case (see e.g., the references in Sec.~\ref{sec:isco-nr-compare} above), several papers (e.g., \cite{blanchet-isco-PRD2002,damour-gourgoulhon-grandclement-ISCO-PRD2002}) take the viewpoint that the ``correct'' quantity to compare with comparable-mass QCID simulations is the ICO rather than the ISCO (although the QCID papers \cite{cook-effectivepotential-PRD1994,baumgarte-isco-PRD2000,pfeiffer-saul-cook-quascirc-PRD2000,gourgoulhon-etal-hkvII-PRD2002,baker-quasicircdata-grqc-2002,cook-pfeiffer-PRD2004,yo-cook-shapiro-baumgarte-PRD2004,tichy-brugmann-initialdata-PRD2004,hannam-quasicirc-PRD2005,alcubierre-etal-quasicirc-PRD2005,caudill-etal-initialdata-PRD2006} themselves usually refer to this quantity as an ISCO). This is in part because an ISCO (in \emph{some} approaches) is not always defined at each PN order or for $\eta \sim 1/4$ (see, e.g., Sec.~\ref{sec:PNiscos} below or Sec.~\mbox{IV~A~2} of \cite{bcv1,*bcv1-erratum}). However, in the small-$\eta$ limit there are both ICO and ISCO methods that yield well-defined (but not always well-behaved) results. This will be made clear in Sec.~\ref{sec:results} below.
In the rest of this paper, for simplicity I will refer to both ICOs and ISCOs as ``ISCOs.'' The context will make clear if the method in question is formally an ICO or an ISCO.

\subsection{\label{sec:GSFreview}A (very) brief overview of the gravitational self-force}
How does the ISCO change when the mass $m_1$ is no longer completely negligible? The answer to this question is the purview of self-force calculations. The self-force causes the motion of the point-mass $m_1$ to deviate from geodesic motion via
\be
\label{eq:selfforceeqn1}
\frac{d^2x^{\alpha}}{d\tau^2} + {\Gamma^{\alpha}_{\beta \gamma}}[g^0] \frac{dx^{\beta}}{d\tau} \frac{dx^{\gamma}}{d\tau} = a^{\alpha}_{\rm self (1)} + a^{\alpha}_{\rm self (2)} + O(q^3),
\ee
where on the right-hand-side the self-force is expanded in powers of the mass ratio [$a^{\alpha}_{\rm self (n)} \propto q^n$]. The background metric $g^0_{\mu \nu}$ used in the left-hand-side is usually taken to be the Schwarzschild or Kerr metric. The full spacetime metric includes perturbative corrections of the form
\be
g_{\mu \nu} = g^0_{\mu \nu} + h^{(1)}_{\mu \nu} + h^{(2)}_{\mu \nu} + O(q^3),
\ee
where $h^{(n)}_{\mu \nu} \propto q^n$.

The leading-order GSF has been derived by several authors (see \cite{poissonselforcereview,lousto-selfforcereview,detweiler-selfforcelec,barack-selfforcereview,gralla-wald-GSFderivationCQG2008,pound-GSFderivationPRD2010} for references and recent work) and is given by
\be
a^{\alpha}_{\rm self (1)}(z) = \bar{\nabla}^{\alpha \beta \gamma} \bar{h}^{\rm R}_{\beta \gamma}(z),
\ee
where $\bar{\nabla}^{\alpha \beta \gamma}$ is a differential operator proportional to a covariant derivative, and $\bar{h}^{\rm R}_{\beta \gamma}(z)$ is the regularized metric perturbation evaluated at the position $z^{\mu}$ of $m_1$ (an overbar means to take the trace-reversed part). This regularized metric perturbation is a smooth solution of the homogenous linearized perturbation equations. It is constructed from the first-order retarded metric perturbation $\bar{h}^{\rm ret}_{\beta \gamma}$ by analytically subtracting out a singular contribution $\bar{h}^{\rm S}_{\beta \gamma}$. The retarded metric perturbation is itself a numeric solution of the inhomogeneous linearized perturbation equations with a point-particle source. Note that the motion of $m_1$ is equivalently described by purely geodesic motion, but in terms of a background metric $g^0_{\mu \nu} + h^{\rm R}_{\mu \nu}$ and a new proper time $\tau'$,
\be
\frac{d^2x^{\alpha}}{d\tau'^2} + {\Gamma^{\alpha}_{\beta \gamma}}[g^0+h^{\rm R}] \frac{dx^{\beta}}{d\tau'} \frac{dx^{\gamma}}{d\tau'} = 0.
\ee
For further details see \cite{barack-selfforcereview,detweiler-selfforcelec} and references therein.

Self-force effects are more easily studied by splitting the GSF into a dissipative (time-odd) and conservative (time-even) piece. If one views the GSF as moving the particle $m_1$ along a sequence of geodesics instantaneously tangent to its motion, then the dissipative and conservative pieces of the GSF modify the constants of motion parametrizing these geodesics. The dissipative GSF piece is responsible for secular changes in the ``intrinsic'' constants of the motion: the energy $E$, azimuthal angular momentum $L$, and the Carter constant $Q$. These changes give rise to radiation-reaction, causing the orbital separation, eccentricity, and inclination to slowly evolve on a radiation-reaction time scale. The conservative GSF also modifies the intrinsic constants of the motion, but does so in an oscillatory manner that averages to zero on an orbital time scale. However, the conservative GSF can also affect the ``extrinsic'' constants of the motion: these constants are responsible for the orientation of the geodesic and the location of the particle on the geodesic.

\subsection{\label{sec:dissiscoreview}Dissipative self-force effects on the ISCO}
The effect of the dissipative GSF on the ISCO was addressed in a study by Ori and Thorne \cite{ot00}. Specifically, they showed that the region near the ISCO gets ``blurred'' into a \emph{transition regime} lying between the adiabatic inspiral and plunging phases. They derive approximate equations of motion for the transition regime by expanding the geodesic equations about small deviations from the ISCO. Dissipative GSF effects are included by allowing for dissipative changes in the $\tilde{E}$ and $\tilde{L}$ that enter the effective potential $V_{\rm eff}$ (note that for Kerr, $V_{\rm eff}$ depends on both $\tilde{L}$ and $\tilde{E}$). They find that the radius of the transition regime $\Delta r=r_{\rm isco}-r_{\rm trans}$ and the shift in the orbital frequency $\Delta \Omega=\Omega_{\rm trans}-\Omega_{\rm isco}$ is given by (Sec.~IV of \cite{ot00})
\be
\label{eq:ISCOshift-diss}
\frac{\Delta r}{m_2} \approx 18 q^{2/5} \;\;\;\text{and} \;\;\;
\frac{\Delta \Omega}{\Omega_{\rm isco}} \approx 4.4 q^{2/5}.
\ee
Note also that the energy and angular momentum radiated during the transition regime is given by [cf.~Eq.~(3.26) of \cite{ot00}]
\be
\label{eq:DeltaEdiss}
\frac{\Delta E}{m_2} \approx 0.096 q^{9/5} \;\;\;\;\;\; \text{and} \;\;\;\;\;\;
\frac{\Delta L}{m_2^2} \approx 1.4 q^{9/5},
\ee
with the corresponding fluxes at infinity given by
\be
\label{eq:DeltaEdtdiss}
\frac{\Delta E}{\Delta t} \approx 0.000\,97 q^2 \;\;\;\;\;\; \text{and} \;\;\;\;\;\;
\frac{\Delta L/m_2}{\Delta t} \approx 0.014 q^2.
\ee
In Eqs.~\eqref{eq:ISCOshift-diss}--\eqref{eq:DeltaEdtdiss}, the numerical prefactors assumed the Schwarzschild spacetime; the corresponding values for equatorial orbits in Kerr are easily derived from \cite{ot00}.
These scalings for the transition region were also independently derived in the EOB analysis of Buonanno and Damour \cite{EOB-BD2} (which considered the nonspinning case but for arbitrary mass ratios).

\subsection{\label{sec:conserviscoreview}The Barack-Sago conservative GSF ISCO shift}
To compute the conservative GSF corrections to the ISCO, Barack and Sago \cite{barack-sago_isco,barack-sago-eccentricselfforce} analyzed the equations of motion in the form
\bs
\begin{align}
\frac{d^2\hat{r}}{d\hat{\tau}^2} &= -\frac{1}{2}\frac{\partial V_{\rm eff}(\hat{r},\hat{\tilde{L}})}{\partial \hat{r}} + a^r_{\rm cons}\\
\frac{d\hat{\tilde{E}}}{d\hat{\tau}} &= a_t^{\rm cons} \;\;\;\;\;\;\;\;\; \frac{d\hat{\tilde{L}}}{d\hat{\tau}} = a_{\varphi}^{\rm cons}
\end{align}
\es
where the hats refer to quantities along the new orbit of the particle (which is no longer a geodesic and on which the energy and angular momentum parameters $\hat{\tilde{E}}$ and $\hat{\tilde{L}}$ are no longer constants of the motion). These equations are then expanded in terms of a slightly eccentric orbit near the Schwarzschild ISCO. The resulting shifts in the ISCO radius and orbital frequency are then expressed in terms of the components of the GSF evaluated at the ISCO. The difficult part of the BS analysis is numerically computing the components of the GSF along circular and eccentric geodesics (see \cite{barack-sago-circselfforcePRD2007,barack-sago_isco,barack-sago-eccentricselfforce} for the details).
Working in Lorenz gauge, BS find\footnote{Recall that most of the self-force literature uses $(\mu,M)$ in place of $(m_1,m_2)$. Because we compare with PN computations, we use here the conventions commonly employed in the PN literature: $M=m_1+m_2$, $\mu=m_1 m_2/M$, $\eta=\mu/M$, and $q=m_1/m_2 \leq 1$.} \cite{barack-sago_isco,barack-sago-eccentricselfforce}
\bs
\be
r_{\rm isco}= 6m_2 - 3.269(\pm 0.002) m_1,
\ee
\be
m_2 \Omega_{\rm isco}^{\rm Lorenz}= 6^{-3/2} [ 1+ 0.4869(\pm 0.0004) q ].
\ee
\es

As pointed out by Damour \cite{damour-GSF} (see also Sec.~\mbox{III~D} of \cite{barack-lousto-Schwpert-PRD2005} and Sec.~\mbox{III~B} of \cite{sago-barack-detweiler}), the Lorenz gauge used for the calculations in BS is not asymptotically flat. Rather the asymptotically-flat time coordinate is related to the $r\rightarrow \infty$ limit of the perturbed binary metric by $dt_{\rm flat} = (1+2\alpha)^{1/2} dt_{\rm Lorenz}$, where
\be
\alpha \equiv \frac{m_1 \hat{\tilde{E}}_1}{r_0 (1-2m_2/r_0)}.
\ee
Here $r_0$ refers to the Schwarzschild coordinate radius of the particle's ($m_1$) circular orbit, and
\be
\hat{\tilde{E}}_1\equiv \frac{E_1}{m_1} = \frac{1-2m_2/r_0}{\sqrt{1-3m_2/r_0}}
\ee
is the particle's conserved energy per unit rest mass. Evaluated at the ISCO ($r_0=6m_2$), $\alpha$ takes the value $q\sqrt{2}/6$. To convert angular circular-orbit frequencies from Lorenz coordinates ($\Omega_{\rm Lorenz}\equiv d\varphi/dt_{\rm Lorenz}$) to asymptotically-flat coordinates ($\Omega_{\rm flat}\equiv d\varphi/dt_{\rm flat}$), we use \cite{damour-GSF}
\be
\Omega_{\rm flat} = \frac{\Omega_{\rm Lorenz}}{\sqrt{1+2\alpha}} \approx \Omega_{\rm Lorenz} [1-\alpha + O(q^2)].
\ee
When converted to asymptotically-flat coordinates, the frequency of the ISCO becomes
\be
\label{eq:Omega_flat}
m_2 \Omega_{\rm flat}^{\rm isco, sf} = 6^{-3/2} \left[1+ \left(c_{\Omega}^{\rm BS} -\frac{\sqrt{2}}{6} \right) q + O(q^2) \right],
\ee
where $c_{\Omega}^{\rm BS} = 0.4869(\pm 0.0004)$ \cite{barack-sago-eccentricselfforce}. To simplify comparisons with PN ISCO calculations, it is convenient to express this result in terms of $M$ and $\eta$ instead of $m_2$ and $q$. Multiplying both sides of Eq.~\eqref{eq:Omega_flat} by $M/m_2$ and using $\eta = q + O(q^2)$, we have
\be
\label{eq:Omega_flat_renorm}
M \Omega_{\rm flat}^{\rm isco, sf} = 6^{-3/2} \left[1+ c_{\Omega}^{\rm ren} \eta + O(\eta^2) \right],
\ee
where
\be
\label{eq:cOmegaren}
c_{\Omega}^{\rm ren}=1+ c_{\Omega}^{\rm BS} -\frac{1}{\sqrt{18}} = 1.2512(\pm 0.0004).
\ee
It is this number that will be compared with the PN-based ISCO calculations below.

Note that the ratio of the conservative to the dissipative [Eq.~\eqref{eq:ISCOshift-diss}] changes to the ISCO radius or frequency is roughly equal to $\sim 0.1 q^{3/5}$. For small mass ratios ($q\sim 10^{-2} \mbox{--} 10^{-7}$) this implies that the ``blurring'' of the ISCO by the dissipative GSF overwhelms the small conservative-GSF shift in the ISCO by a factor of $\sim 10^2 \mbox{--} 10^5$. While the dissipative GSF ISCO shift becomes more important than the conservative shift as the mass ratio gets smaller, in the comparable-mass limit the entire notion of an ISCO is not well-defined (at least in the presence of dissipation). The conservative-GSF ISCO-shift is therefore unlikely to be a quantity of observational importance. Rather its importance lies in the fact that it represents a unique, strong-field critical point in the conservative two-body dynamics that can serve as a test of numeric GSF codes and a point of comparison with PN (and perhaps NR) calculations.
\section{\label{sec:PNiscos}A catalog of PN-based methods for computing the ISCO}
This section reviews nearly all PN-based methods for computing the ISCO. For each method discussed below, a \textsc{Maple} code was developed to numerically compute the ISCO frequency $\Omega_{\rm isco}(\eta)$ as a function of the reduced mass ratio.
\subsection{\label{sec:pnenergy}PN energy function}
One of the simplest methods for determining the ISCO (in this case an ICO) is to minimize the PN circular-orbit energy $E^{\rm PN}(\Omega)$ with respect to frequency. For nonspinning binaries, this energy is given by [see Eq.~(3) of \cite{blanchet-isco-PRD2002} and references therein\footnote{For extensions to the case of spinning binaries, see \cite{kidderwillwiseman-spineffects,kidder-spineffects,damour-gourgoulhon-grandclement-ISCO-PRD2002,blanchet-isco-PRD2002,faye-buonanno-luc-higherorderspinI,faye-buonanno-luc-higherorderspinII,*faye-buonanno-luc-higherorderspinIIerratum,*faye-buonanno-luc-higherorderspinIIerratum2}. For eccentric or tidally distorted binaries, see \cite{mora-will-PRD2002,mora-will-PRD2004,*mora-will-PRD2004-erratum,berti-iyer-will-PRD2006,berti-iyer-will-PRD2008}.}]
\begin{multline}
\label{eq:E3PN}
\frac{E^{\rm PN}(\Omega)}{\eta M} = -\frac{1}{2} x \left\{ 1 + x \left(-\frac{3}{4} - \frac{\eta}{12}\right)
\right. \\
+ x^2 \left( -\frac{27}{8} +\frac{19}{8}\eta - \frac{\eta^2}{24}\right) + x^3 \left[ -\frac{675}{64}  \right.
\\ \left.
+ \left( \frac{34\,445}{576} -\frac{205}{96}\pi^2 \right) \eta
- \frac{155}{96}\eta^2 - \frac{35}{5184}\eta^3\right]
\\
+ x^4 \left( -\frac{3969}{128} + \eta e_4(\eta) + \frac{448}{15}\eta \ln x \right)
\\ \left.
+ x^5 \! \left( -\frac{45\,927}{512} + \eta e_5(\eta) + \! \left[ -\frac{4988}{35} - \frac{1904}{15} \eta \right] \eta \ln x \right)\! \right\},
\end{multline}
where $x\equiv (M\Omega)^{2/3}$ and the known value for the regularization parameter $\lambda=-1987/3080$ \cite{damour-jaranowski-schafer-dimreg-PhLettB2001,blanchet-damour-farese-dimreg,itoh-futamase-3PNreg-PRD2003,itoh-3PNreg-PRD2004,blanchetdamour3PNprl} is used throughout this article. Equation \eqref{eq:E3PN} also includes newly computed logarithmic correction terms at 4PN and 5PN orders \cite{blanchet-etal-selfforceII}. The test-mass pieces of these 4PN and 5PN terms are known from the exact Schwarzschild expression [Eq.~\eqref{eq:Eschw} below]. The functions $e_4(\eta)$ and $e_5(\eta)$ denote some unknown polynomials in $\eta$. Section \ref{sec:fitEpn} discusses how we can use our present knowledge of the ISCO from GSF and QCID calculations to help constrain these functions. However, the ISCO comparisons discussed in Secs.~\ref{sec:results} and \ref{sec:discuss} below will only make use of the circular-orbit energy to 3PN order.

Note that in the small-mass-ratio limit, it is well-known that standard Taylor PN expansions converge slowly \cite{cutler-etal-last3minutesPRL1993,poisson-bhpertVI-PNaccuracy,*poisson-bhpertVI-PNaccuracy-erratum,simone-leonard-poisson-will-CQG1997,leonard-poisson-CQG1998,yunes-berti-PNaccuracy,blanchet-PNaccuracy-confproc}.
For example, Taylor expanding the Schwarzschild circular-orbit energy
\be
\label{eq:Eschw}
\frac{E^{\rm Schw}(\Omega)}{\eta M} = \frac{(1-2x)}{(1-3x)^{1/2}} -1,
\ee
and computing the ISCO frequency at each PN order, one needs to go to at least 4PN order to reproduce the exact test-mass ISCO ($M \Omega_{\rm Schw}=6^{-3/2}\approx 0.0680$) to within $12\%$ (see also Sec.~II of Ref.~\cite{damour-jaranowski-schafer-EOBisco}). Since the (nonresummed) 3PN energy function poorly matches the test-mass ISCO frequency [Eq.~\eqref{eq:E3PN} predicts $M\Omega^{\rm 3PN}_{\rm isco} (\eta=0)\approx 0.0867$], this method cannot be straightforwardly compared with the BS conservative GSF ISCO shift.
\subsection{\label{sec:pneqns}Stability analysis of the PN equations of motion}
The ISCO can be determined by directly analyzing the conservative-pieces of the PN two-body equations of motion. In harmonic coordinates the relative two-body acceleration can be written in the form
\be
\label{eq:pngeneral}
{\bm a} = -(M/r^2) [ A(r,\dot{r},\dot{\varphi}) {\bm n} + B(r,\dot{r},\dot{\varphi}) {\bm v}],
\ee
where $M=m_1+m_2$ is the total mass, $r$ is the relative orbital separation, ${\bm n}$ is a unit vector that points along the relative separation vector, and ${\bm v}$ is the relative orbital velocity. The orbital phase angle is denoted $\varphi$ and an overdot refers to a derivative with respect to coordinate time $t$. The functions $A=1+A^{\rm 1PN}+A^{\rm 2PN} + A^{\rm 2.5PN} + A^{\rm 3PN} + A^{\rm 3.5PN}$ and $B=B^{\rm 1PN}+B^{\rm 2PN} + B^{\rm 2.5PN} + B^{\rm 3PN} + B^{\rm 3.5PN}$ are known to 3.5PN order (see \cite{blanchetLRR} for references). For illustration, the 1PN pieces are
\bs
\label{eq:1PNconservpieces}
\begin{align}
A^{\rm 1PN} &= -2(2+\eta)\frac{M}{r} +(1+3\eta)v^2 -\frac{3}{2} \eta \dot{r}^2, \\
B^{\rm 1PN} &= -2 (2-\eta)\dot{r},
\end{align}
\es
where $v^2=\dot{r}^2+r^2\dot{\varphi}^2$ for planar motion.
The remaining pieces can be found in Eqs.~(181)\mbox{--}(186) of \cite{blanchetLRR}. Note that at 3PN order I use the form given in Eqs.~(185)\mbox{--}(186) in which a gauge transformation has been applied to remove the logarithmic terms. Since we are concerned only with conservative effects, the radiation-reaction pieces at 2.5PN and 3.5PN orders are set to zero.

To compute the ISCO we follow the prescription given in Sec.~\mbox{III~A} of Ref.~\cite{kidderwillwiseman-transition-PRD1993}. This involves (i) expressing Eq.~\eqref{eq:pngeneral} as a system of first-order equations for the variables $r$, $\omega\equiv \dot{\varphi}$, $u\equiv \dot{r}$, (ii) linearizing those equations about a circular-orbit solution ($\dot{r}=\dot{u}=\dot{\omega}=0$), and (iii) finding a criteria for the stability of the solution. This produces a system of equations for the ISCO radius and orbital frequency $(r_0, \omega_0)$ [Eqs.~(3.3) and (3.6) of \cite{kidderwillwiseman-transition-PRD1993}]:
\bs
\label{eq:ISCOcondKWW}
\begin{align}
\omega_0^2 &= M A_0/r_0^3,\\
\label{eq:kwwcondition}
C_0 &\equiv 1-2\frac{\omega_0}{A_0} \left(\frac{\partial A}{\partial \omega}\right)_0 + \frac{r_0}{A_0}\left(\frac{\partial A}{\partial r}\right)_0 \nonumber
\\
& \;\;\;\; + 2\frac{M}{r_0} \left(\frac{\partial B}{\partial u}\right)_0 \left[ 1-\frac{1}{2}\frac{\omega_0}{A_0} \left(\frac{\partial A}{\partial \omega}\right)_0 \right] =0,
\end{align}
\es
where the subscript $0$ means to evaluate the quantities along a circular orbit.

Solving these equations numerically yields well-defined solutions at 2PN order, but no physical solutions at 1PN or 3PN orders. Even at 2PN order, in the test mass limit the ISCO is found to be at $r=6.505 M$ (see Table II of \cite{kidderwillwiseman-transition-PRD1993}) instead of the correct harmonic coordinate radius of $5M$. Strangely, as the reduced mass ratio is increased from $0$ to $1/4$, both the ISCO radius and the ISCO angular frequency increase. This approach does not appear to yield sensible results for the ISCO.

Blanchet and Iyer \cite{blanchetiyer3PN} pursued a variation of the above approach (see also \cite{blanchet-BIisco-confproc}). Instead of solving Eqs.~\eqref{eq:ISCOcondKWW} numerically, they derived a PN series expansion for $\omega_0$ and $\hat{C}_0$ [a quantity equivalent to Eq.~\eqref{eq:kwwcondition}] in terms of the harmonic radial coordinate\footnote{Note that in their derivation of Eqs.~\eqref{eq:BIiscocond1}, Ref.~\cite{blanchetiyer3PN} used the 3PN equations of motion in a gauge in which the logarithmic terms are still present. These logarithmic terms depend on an arbitrary gauge constant $r_0'$ associated with the choice of coordinates. Also note that we introduce here the notation $O(n)$ to refer to terms of order $n/2$ PN [i.e., $O(c^{-n})$].},
\bs
\label{eq:BIiscocond1}
\begin{multline}
\label{eq:omega0_harmonic}
\omega_{0, \rm har.}^2 = \frac{M}{r_0^3} \left\{ 1+ \frac{M}{r_0}(-3+\eta) + \frac{M^2}{r_0^2} \left(6+\frac{41}{4}\eta + \eta^2\right)
\right. \\
+ \frac{M^3}{r_0^3} \left( -10 + \left[\frac{-75\,707}{840} + \frac{41}{64}\pi^2 + 22\ln\left(\frac{r_0}{r_0'}\right)\right] \eta
\right. \\ \left. \left.
+ \frac{19}{2}\eta^2 +\eta^3 \right) + O(8)\right\},
\end{multline}
\begin{multline}
\hat{C}_0^{\rm har.} = \frac{M}{r_0^3} \left\{ 1+ \frac{M}{r_0}(-9+\eta) + \frac{M^2}{r_0^2} \left(30+\frac{65}{4}\eta + \eta^2\right)
\right. \\
+ \frac{M^3}{r_0^3} \left( -70 + \left[\frac{-29\,927}{840} - \frac{451}{64}\pi^2 + 22\ln\left(\frac{r_0}{r_0'}\right)\right] \eta
\right. \\ \left. \left.
+ \frac{19}{2}\eta^2 +\eta^3 \right) +O(8) \right\}=0 .
\end{multline}
\es
Reference \cite{blanchetiyer3PN} also derived relationships equivalent to Eqs.~\eqref{eq:BIiscocond1} but starting from the 3PN Hamiltonian in ADM coordinates [see their Eqs.~(6.24) and (6.38)]. These Hamiltonian-based expressions $\omega_{0, \rm ADM}^2$ and $\hat{C}_0^{\rm ADM}$ are expressed in terms of the ADM radial coordinate $R$ and differ from Eqs.~\eqref{eq:BIiscocond1} at 2PN and higher orders. However, Ref.~\cite{blanchetiyer3PN} showed that the two sets of equations agree if one applies the coordinate transformation between the ADM and harmonic coordinate radii. In either formulation one can solve the equation for $\hat{C}_0$ to determine the ISCO radius and substitute the result into the equation for $\omega_0^2$ to find the corresponding orbital frequency. However, because of the coordinate-dependent nature of these two formulations, they yield different numerical results at 3PN order\footnote{At 1PN order both methods are identical; at 2PN order there is no ISCO in either formulation.}: in the test-mass limit $r_0^{\rm 3PN} \approx 5.93 M$ ($M\omega_{0, {\rm har.}}^{\rm 3PN} \approx 0.0544$) while in the Hamiltonian formulation $R_0^{\rm 3PN} \approx 5.76 M$ ($M\omega_{0, {\rm ADM}}^{\rm 3PN}\approx 0.0561$). This $\approx 3\%$ difference in the frequencies presumably arises from differences at 4PN and higher orders.
\subsubsection{\label{sec:C0invar}Gauge invariant description of the PN ISCO}
Blanchet and Iyer \cite{blanchetiyer3PN} also derived a gauge-invariant form for the ISCO condition $\hat{C}_0=0$. This comes from
solving Eq.~\eqref{eq:omega0_harmonic} for $r_0$ (or $R_0$ in the ADM case) in terms of the PN frequency parameter $x_0\equiv(M\omega_0)^{2/3}$, and substituting the resulting expression for $r_0(x)$ [or $R_0(x)$] into the expression for $\hat{C}_0^{\rm har.}$ (or $\hat{C}_0^{\rm ADM}$). The result is an expression for $\hat{C}_0 \equiv C_0 M^2/x_0^3$ [their Eq.~(6.1)] that depends directly on the ISCO orbital frequency and not on any gauge-dependent radius,
\begin{multline}
\label{eq:C0invar}
\hat{C}_0 = 1-6x_0 + 14\eta x_0^2
\\
+ \left[ \left(\frac{397}{2} -\frac{123}{16}\pi^2\right)\eta -14\eta^2\right]x_0^3 + O(8).
\end{multline}
This expression has the very interesting property that it yields the exact test-mass Schwarzschild result ($x^{\rm isco}=1/6$) at all PN orders \emph{without any form of ``resummation.''} Finite-mass-ratio effects do not enter until 2PN order. In this case the ISCO exists for all $\eta$ and can be solved analytically to yield
\begin{align}
M\omega_{\rm ISCO}^{\rm 2PN} &= \left[\left(\frac{3}{14\eta}\right)\left(1-\sqrt{1-\frac{14}{9}\eta}\right) \right]^{3/2} \\
& = 6^{-3/2} [1+7\eta/12 +O(\eta^2)].
\end{align}
At 3PN order the ISCO only exists for $\eta<0.183$; for larger mass ratios all circular orbits are stable. Equation \eqref{eq:C0invar} can be solved exactly at 3PN order, but not in a simple form. Its expansion for small $\eta$ is given by
\bs
\label{eq:3PNC0iscofreq}
\be
x_{\rm ISCO}^{\rm 3PN} = 6^{-1} \left[ 1 + c_x^{C_0{\rm 3PN}} \eta + O(\eta^2) \right], \;\;\text{or}
\ee
\be
M\omega_{\rm ISCO}^{\rm 3PN} = 6^{-3/2} \left[ 1 + c_{\omega}^{C_0{\rm 3PN}} \eta + O(\eta^2) \right],
\ee
\es
where
\begin{align}
\label{eq:c3PNvalues}
c_x^{C_0{\rm 3PN}} &\equiv \frac{565}{432}-\frac{41\pi^2}{1152} = 0.956\,608\,407\ldots, \\
c_{\omega}^{C_0{\rm 3PN}} &\equiv \frac{565}{288}-\frac{41\pi^2}{768} = 1.434\,912\,612\ldots.
\end{align}
Note, in particular, that the coefficient $c_{\omega}^{C_0{\rm 3PN}}\approx 1.435$ differs from the exact value $c_{\Omega}^{\rm ren}=1.251$ by $14.7\%$. As we will see in Secs.~\ref{sec:results} and \ref{sec:discuss} below, this agreement is better than any 3PN order estimate, including all hybrid, resummed, or EOB methods. The extension of the above ISCO condition to spinning systems is discussed in \cite{favata-PNspinisco} and briefly in Sec.~\ref{sec:discuss} below.
\subsection{\label{sec:pnhamiltonin}PN Hamiltonian}
While Ref.~\cite{blanchetiyer3PN} determined an analytic condition for the ISCO using the PN Hamiltonian in ADM coordinates, an alternative method follows from the work of Ref.~\cite{wex-schafer-CQG1993}. Here one starts with the reduced PN Hamiltonian,
\be
\label{eq:ADMham}
\hat{\mathcal H} \equiv \frac{\mathcal H^{\rm ADM}}{\mu} = \hat{\mathcal H}_{\rm N} + \hat{\mathcal H}_{\rm 1PN} + \hat{\mathcal H}_{\rm 2PN} +\hat{\mathcal H}_{\rm 3PN},
\ee
where the Newtonian and 1PN terms are
\bs
\be
\hat{\mathcal H}_{\rm N} = \frac{\hat{\bm P}^2}{2} - \frac{1}{\hat{R}} \;\;\; \text{and}
\ee
\begin{multline}
\hat{\mathcal H}_{\rm 1PN} = - \frac{1}{8}(1-3\eta) \hat{\bm P}^4
\\
- \frac{1}{2}\frac{1}{\hat{R}} [(3+\eta)\hat{\bm P}^2 + \eta ({\bm N}\cdot \hat{\bm P})^2] + \frac{1}{2}\frac{1}{\hat{R}^2},
\end{multline}
\es
$R\equiv \hat{R} M$ is the ADM radial coordinate, $\hat{{\bm P}}$ is the conjugate momenta divided by the reduced mass $\mu$, and ${\bm N}\equiv {\bm R}/R$ is the unit orbital separation vector. The 2PN and 3PN terms can be read from Eq.~(5.9) of \cite{blanchetiyer3PN} or references therein. Hamilton's equations are
\begin{align}
\frac{d{R}}{d{t}} &= \frac{\partial {{\mathcal H}^{\rm ADM}}}{\partial {P}_R},& \frac{d\Psi}{d{t}} &= \frac{\partial {{\mathcal H}^{\rm ADM}}}{\partial {P}_{\Psi}},\\
\frac{d{P}_R}{d{t}} &= -\frac{\partial {{\mathcal H}^{\rm ADM}}}{\partial {R}},& \frac{d{P}_{\Psi}}{d{t}} &= 0,
\end{align}
where $(R,\Psi)$ are ADM polar coordinates.

The conditions for the ISCO are
\be
P_R=0, \;\; \dot{P}_R=-\frac{\partial {\mathcal H}^{\rm ADM}}{\partial R}=0, \;\; \frac{\partial \dot{P_R}}{\partial R} = -\frac{\partial^2 {\mathcal H}^{\rm ADM}}{\partial R^2} =0.
\ee
Following \cite{wex-schafer-CQG1993}, the conserved energy $E=\mu \hat{E}$ and angular momentum ${\bm J} = {\bm P}_{\Psi}=\hat{{\bm J}} \mu M$ are
\be
\hat{E}=\hat{\mathcal H}({\hat {\bm R}},{\hat {\bm P}}) \;\;\;\;\;\;\;\;\; \hat{{\bm J}} = \hat{\bm R} \times \hat{\bm P},
\ee
and the above conditions for the ISCO are equivalent to the system
\be
\label{eq:EhamISCOcond}
\frac{\partial \hat{E_0}}{\partial \hat{R}}(\hat{R},\hat{J}) =0 \;\;\;\;\;\;\;\;\; \frac{\partial^2 \hat{E_0}}{\partial \hat{R}^2}(\hat{R},\hat{J}) =0,
\ee
where $\hat{E}_0(\hat{R},\hat{J})$ is the energy evaluated along a circular orbit and is obtained by substituting ${\bm N}\cdot \hat{\bm P}=0$ and $\hat{\bm P}=\hat{J}^2/\hat{R}^2$ (with $\hat{J}=|\hat{\bm J}|$) into Eq.~\eqref{eq:ADMham} [see Eq.~(5) of \cite{wex-schafer-CQG1993}]. For example, up to 1PN order the energy is \cite{wex-schafer-CQG1993}
\begin{multline}
\label{eq:E_PNham}
\hat{E}_0(\hat{R},\hat{J}) = \frac{1}{2} \frac{\hat{J}^2}{\hat{R}^2} - \frac{1}{\hat{R}} - \frac{1}{8} \left(\frac{\hat{J}^4}{\hat{R}^4} + 12 \frac{\hat{J}^2}{\hat{R}^3} - \frac{4}{\hat{R}^2} \right)+ O(4)
\\
+ \eta \left[ \frac{1}{8} \left(3\frac{\hat{J}^4}{\hat{R}^4} - 4 \frac{\hat{J}^2}{\hat{R}^3}\right) + O(4) \right],
\end{multline}
where the $\eta$-dependent terms have been separated to illustrate the hybrid method discussed below.

By solving Eqs.~\eqref{eq:EhamISCOcond} numerically the ISCO values $\hat{R}_0$ and $\hat{J}_0$ are determined. The corresponding ISCO frequency is found from
\be
\label{eq:Omegaisco_pnham}
M\Omega_0 = \left. M \frac{d\Psi}{dt}\right|_0 = \left. \frac{\partial \hat{E}}{\partial \hat{J}}\right|_0.
\ee
At 2PN order no solutions for the ISCO were found. At 1PN order there are solutions, although in the test-mass limit they differ significantly ($M\Omega^{\rm 1PN}_0\approx 0.0234$, $\hat{R}_0^{\rm 2PN}=11.1$) from the exact result (as expected). At 3PN order the results are somewhat better ($M\Omega^{\rm 3PN}_0\approx 0.0396$, $\hat{R}_0^{\rm 3PN}=7.55$) but still differ by $42\%$ from the exact test-mass frequency. This method does not appear to be well-suited to finding the ISCO.
\subsection{\label{sec:hybridmethods}Hybrid methods}
With the exception of the gauge-invariant method discussed in Sec.~\ref{sec:C0invar}, the above methods cannot reproduce the correct ISCO frequency in the test mass limit. This is not necessarily surprising since (i) the ISCO occurs in the strong field where the PN expansion starts to break down, and (ii) the PN expansion is known to converge more slowly in the test-mass limit. To help remedy this problem, a hybrid approach was introduced by Kidder, Will, and Wiseman \cite{kidderwillwiseman-transition-CQG1992,kidderwillwiseman-transition-PRD1993}  to enforce the standard Schwarzschild dynamics in the test-mass limit. The basic philosophy behind the hybrid approach is to replace the test-mass limit parts of some PN expanded function (i.e., the leading-order terms in an expansion in $\eta$) with the equivalent terms from the exact Schwarzschild representation of the same function. Many of the previously discussed methods for computing the ISCO have hybrid analogs which we now describe.
\subsubsection{\label{sec:hybridenergy}Hybrid energy function}
A hybrid version of the 3PN energy function in Eq.~\eqref{eq:E3PN} is easily computed by removing the test-mass pieces and replacing them with
Eq.~\eqref{eq:Eschw}. [Note that the PN expansion of Eq.~\eqref{eq:Eschw} coincides with the test-mass limit of Eq.~\eqref{eq:E3PN}.] The result is
\begin{multline}
\label{eq:Ehybrid}
\frac{E_{\rm hybrid}^{\rm PN}}{\eta M} = \frac{(1-2x)}{(1-3x)^{1/2}} -1
\\
-\frac{x}{2} \left\{
- \frac{\eta}{12} x + x^2 \left( \frac{19}{8}\eta - \frac{\eta^2}{24}\right)
\right. \\ \left.
+ x^3 \left[ \left( \frac{34\,445}{576} -\frac{205}{96}\pi^2 \right) \eta - \frac{155}{96}\eta^2 - \frac{35}{5184}\eta^3\right]
 \right\}.
\end{multline}
This is equivalent to expressions found in Refs.~\cite{ajith-iyer-robinson-sathya_PNapproximants,*ajith-iyer-robinson-sathya_PNapproximants_errata,porter-PRD2007} (which were inspired by the approach in \cite{kidderwillwiseman-transition-CQG1992,kidderwillwiseman-transition-PRD1993}). Unlike some of the other methods investigated here, this hybrid energy function produces an ISCO (more appropriately an ICO) that is uniquely defined at all PN orders and for all values of $\eta$. Like all of the remaining hybrid, resummation, and EOB methods discussed below, it reproduces the exact Schwarzschild result for the ISCO in the test-mass limit.
\subsubsection{\label{sec:hybridpneqns}Hybrid PN equations of motion}
In the original KWW hybrid approach \cite{kidderwillwiseman-transition-CQG1992,kidderwillwiseman-transition-PRD1993}, the conservative parts of $A$ and $B$ in Eq.~\eqref{eq:pngeneral} were split into test-mass and non-test-mass pieces. The PN test-mass pieces were then replaced by the exact test-mass terms that arise from the geodesic equation of the Schwarzschild metric written in harmonic coordinates (see Sec.~\mbox{II~A} of \cite{kidderwillwiseman-transition-PRD1993}). This defines new (hybrid) equations of motion
\be
\label{eq:hybrideqns}
{\bm a}_H = -(M/r^2) [ (A_{\rm S}+A_{\eta}) {\bm n} + (B_{\rm S}+B_{\eta}) {\bm v}],
\ee
where the Schwarzschild pieces are [Eqs.~(2.6) of \cite{kidderwillwiseman-transition-PRD1993}]
\bs
\begin{align}
A_{\rm S} &= \left[ \frac{1-M/r}{(1+M/r)^3}\right] - \left[ \frac{2-M/r}{1-(M/r)^2}\right] \frac{M}{r} \dot{r}^2 +v^2, \\
B_{\rm S} &= -\left[ \frac{4-2M/r}{1-(M/r)^2}\right] \dot{r},
\end{align}
\es
and the non-test-mass PN pieces $A_{\eta}$ and $B_{\eta}$ are found by removing the $\eta$-independent terms from $A^{\rm 1PN}$, $A^{\rm 2PN}$, $A^{\rm 3PN}$, $B^{\rm 1PN}$, $B^{\rm 2PN}$, and $B^{\rm 3PN}$. (As in Sec.~\ref{sec:pneqns}, we ignore the dissipative terms at 2.5PN and 3.5PN orders.) The ISCO radius and angular orbital frequency is then computed by numerically solving Eqs.~\eqref{eq:ISCOcondKWW} (substituting $A\rightarrow A_{\rm S} + A_{\eta}$ and $B\rightarrow B_{\rm S} + B_{\eta}$). Like all of the hybrid and other remaining methods below, the KWW approach yields the exact Schwarzschild ISCO in the test-mass limit.

This hybrid approach was criticized by Ref.~\cite{wex-schafer-CQG1993}. They developed a Hamiltonian formulation of the hybrid approach (discussed in Sec.~\ref{sec:hybridhamiltonian} below) which gives different results for the ISCO.\footnote{The KWW hybrid approach was also criticized in Ref.~\cite{damour-iyer-sathyaprakash-PRD1998}, which argued that some of the finite-$\eta$ terms in the 2PN equations of motion [i.e., terms in $A_{\eta}$ and $B_{\eta}$] amount to very large fractional corrections to the test-mass terms. However, this argument is not entirely justified. It is more appropriate to compare the ratios $|A_{\eta}/A_{S}|$ or $|B_{\eta}/B_{S}|$, which I find are typically $\lesssim 0.3$ at the ISCO. The Schwarzschild terms do in fact dominate the finite-$\eta$ terms, although not by a large factor. In comparison, the $O(\eta)$ terms in the EOB effective metric potential $A(r)$ [Eq.~\eqref{eq:A-taylor} below] are much smaller than the leading-order Schwarzschild term (by factors $\lesssim 0.009$ near the ISCO) and decrease very quickly with increasing radius. This suggests that the EOB approach is a much better perturbative scheme than the KWW equations.} More troubling, Ref.~\cite{kidderwillwiseman-transition-PRD1993} found that  while at 1PN order the ISCO radius moves inward as $\eta$ is increased, at 2PN order it moves outward (see Fig.~4 of Ref.~\cite{kidderwillwiseman-transition-PRD1993}), in contradiction to the exact result of Ref.~\cite{barack-sago_isco}. Here I extended the KWW hybrid approach to include 3PN order terms. At 1PN and 3PN orders, the ISCO radius moves inward (as expected) for small $\eta$; but unlike at 2PN order, as $\eta$ increases the solutions exhibit a discontinuous jump, after which they move to larger radii. Above some critical value for $\eta$, no solutions for the ISCO are found at 1PN and 3PN orders.
\subsubsection{\label{sec:hybridhamiltonian}Hybrid PN Hamiltonian}
Another hybrid approach (suggested in \cite{wex-schafer-CQG1993}) involves replacing the test-mass pieces of the PN Hamiltonian with the Hamiltonian of the Schwarzschild spacetime in isotropic coordinates. As implemented in \cite{wex-schafer-CQG1993}, this involves replacing the test-mass terms in the first line of Eq.~\eqref{eq:E_PNham} with the reduced circular-orbit energy of a particle in the Schwarzschild spacetime [Eq.~(8) of \cite{wex-schafer-CQG1993}],
\be
\hat{E}_{0, {\rm iso}}^{\rm Schw}(\hat{R},\hat{J}) = \frac{1-1/(2\hat{R})}{1+1/(2\hat{R})} \! \left[1+\left(1+ \frac{1}{2\hat{R}} \right)^{-4} \frac{\hat{J}^2}{\hat{R}^2}  \right]^{1/2} \! -1.
\ee
The resulting energy function is substituted into Eqs.~\eqref{eq:EhamISCOcond} and \eqref{eq:Omegaisco_pnham} to determine the ISCO.

As pointed out in \cite{wex-schafer-CQG1993}, this approach yields different numerical results from \cite{kidderwillwiseman-transition-PRD1993} (even after correcting for the change in coordinate systems). However, it does behave qualitatively similar to the KWW hybrid approach: the ISCO radius moves inward at 1PN and 3PN orders, and outward at 2PN order (and solutions were again not found above some critical $\eta$ at 1PN and 3PN orders).
\subsection{\label{sec:emethod}Resummation approaches: the $e$- and $j$-methods}
Additional methods for computing the ISCO---based on minimizing the Pad\'{e} approximants\footnote{The Pad\'{e} approximant of some function $f$ whose power series is $f(z)=\sum a_n z^n$ consists of a ratio of power series $P_p^q[f(x)]=(\sum^q_i b_i z^i)/(\sum^p_j c_j z^j)$. Pad\'{e} approximants are useful because they tend to accelerate the convergence of a series.} of certain functions---were introduced in Refs.~\cite{damour-iyer-sathyaprakash-PRD1998,damour-jaranowski-schafer-EOBisco}. The \emph{$e$-method} \cite{damour-iyer-sathyaprakash-PRD1998} involves minimizing the Pad\'{e} approximant of a new energy function $e(x)$ defined by
\be
\label{eq:emethod}
e(x)= \left[ \frac{(M+E)^2-m_1^2-m_2^2}{2 m_1 m_2}  \right]^2 -1,
\ee
where $E$ is the orbital energy whose PN expansion for circular binaries is given by Eq.~\eqref{eq:E3PN}. The justification for this expression is discussed in \cite{damour-iyer-sathyaprakash-PRD1998,damour-jaranowski-schafer-EOBisco}. The explicit form of $e(x)$ that is minimized to determine the ISCO is obtained by: (i) substituting the PN expansion $E=E^{\rm PN}$ from Eq.~\eqref{eq:E3PN} into Eq.~\eqref{eq:emethod}, (ii) Taylor expanding the resulting expression for $e(x)$ in terms of $x$ to 2PN order (denoting the result as $T_2{\bm [}e(x){\bm ]}$) or to 3PN order ($T_3{\bm [}e(x){\bm ]}$), and (iii) computing the Pad\'{e} approximant of the result (denoted as $e_{P_2}(x)=P_1^1{\bm [}T_2{\bm [}e(x){\bm ]]}$ or $e_{P_3}(x)=P_2^1{\bm [}T_3{\bm [}e(x){\bm ]]}$). Explicit expressions for $e_{P_2}$ and $e_{P_3}$ are given in Eqs.~(4.6) of \cite{damour-jaranowski-schafer-EOBisco}. They reproduce the Schwarzschild ISCO in the test-mass limit. (See \cite{mroue-kidder-saul-PRD2008} for a criticism of this approach.)

In the \emph{$j$-method}, Ref.~\cite{damour-jaranowski-schafer-EOBisco} proposed another function---$j(x)\equiv J/(\mu M)$, where $J$ is the magnitude of the orbital angular momentum of the system---whose extremum also defines an ISCO. This function is computed by taking the 1PN, 2PN, and 3PN Taylor series for $j^2(x)$ ($T_1{\bm [}j^2(x){\bm ]}$, $T_2{\bm [}j^2(x){\bm ]}$, $T_3{\bm [}j^2(x){\bm ]}$) and constructing the Pad\'{e} approximants $j^2_{P_1}(x) \equiv P_1^0{\bm [}T_1{\bm [}j^2(x){\bm ]]}$, $j^2_{P_2}(x) \equiv P_1^1{\bm [}T_2{\bm [}j^2(x){\bm ]]}$, and $j^2_{P_3}(x) \equiv P_2^1{\bm [}T_3{\bm [}j^2(x){\bm ]]}$. Explicit expressions are found in Eqs.~(4.16) of \cite{damour-jaranowski-schafer-EOBisco}. Reference \cite{damour-jaranowski-schafer-EOBisco} argued that the $j$-method is preferable to the $e$-method because unlike the 1PN Pad\'{e} approximant $e_{P_1}(x) \equiv P_1^0{\bm [}T_1{\bm [}e(x){\bm ]]}$, the test-mass limit of the 1PN Pad\'{e} approximant $j^2_{P_1}(x)$ already reproduces the Schwarzschild ISCO. Additional desirable properties of the $j$-method are discussed in \cite{damour-jaranowski-schafer-EOBisco}.\footnote{Note that \cite{damour-jaranowski-schafer-EOBisco} also defines a third invariant function for computing the ISCO (the \emph{k-method}) that is related to the periastron advance rate. Reference \cite{damour-jaranowski-schafer-EOBisco} considers this method less preferable than the others so I do not consider it here.}
\subsection{\label{sec:EOBmethods}EOB methods}
The effective-one-body (EOB) approach models the conservative two-body dynamics in terms of the dynamics of a single particle in the background of a deformed Schwarzschild geometry. The dissipative dynamics is incorporated by supplementing Hamilton's equations with radiation-reaction forces [these are based on various ways of ``resumming'' the energy flux (see e.g., \cite{damour-iyer-sathyaprakash-PRD1998,damour-iyer-nagar,buonanno-caltechEOB09})].
Since we are concerned with purely conservative corrections to the ISCO, we need only consider the conservative EOB dynamics, which satisfy Hamilton's equations [Eqs.~(2.7)\mbox{--}(2.10) of \cite{EOB-BD2}],
\bs
\label{eq:EOB-conserve}
\begin{align}
\frac{dR}{dT} &= \frac{\partial H^{\rm real}}{\partial P_R}(R,P_R,P_\varphi), \\
\label{eq:EOB-dphidt}
\frac{d\varphi}{dT} &= \frac{\partial H^{\rm real}}{\partial P_{\varphi}}(R,P_R,P_\varphi), \\
\frac{dP_R}{dT} &= -\frac{\partial H^{\rm real}}{\partial R}(R,P_R,P_\varphi), \\
\frac{dP_{\varphi}}{dT} &= 0.
\end{align}
\es
Here motion is restricted to the equatorial plane, and $H^{\rm real}=\mu \hat{H}^{\rm real}$ is the nonspinning real EOB Hamiltonian [Eq.~(2.11) of \cite{EOB-BD2}],
\be
H^{\rm real}(R,P_R, P_{\varphi}) = M \sqrt{1+2\eta (H^{\rm eff}/\mu-1)}.
\ee
The effective EOB Hamiltonian $H^{\rm eff} = \mu \hat{H}^{\rm eff}$ is [e.g., Eq.~(5) of \cite{EOB-damour-nagar-finalspin}]
\be
\hat{H}^{\rm eff} = \sqrt{A(r) \left(1+\frac{p_{\varphi}^2}{r^2} + \frac{p_r^2}{B(r)} + z_3 \frac{p_r^4}{r^2} \right)},
\ee
with $z_3=2\eta(4-3\eta)$, $p_{\varphi}=P_{\varphi}/(\mu M)$, $p_r=P_r/\mu$, $r=R/M$, and $\hat{t}=T/M$.  The functions $A(r)$ and $B(r)$ appear in the EOB effective metric [in Schwarzschild gauge; see, e.g., Eq.~(1) of \cite{EOB-damour-nagar-finalspin}],
\be
ds^2_{\rm eff} = -A(r) d\hat{t}^2 +B(r) dr^2 + r^2 (d\theta^2 + \sin^2\theta d\varphi^2).
\ee

The Taylor expansions of the functions appearing in the EOB metric have been computed to 3PN order and are given by \cite{EOB-BD1,damour-jaranowski-schafer-EOBisco}
\be
\label{eq:A-taylor}
A(r) = 1 -\frac{2}{r} +\frac{2\eta}{r^3} + \frac{a_4(\eta)}{r^4} + \frac{a_5(\eta)}{r^5}, \;\; \text{and }
\ee
\be
B(r) A(r) \equiv D(r) = 1- \frac{6\eta}{r^2} + 2(3\eta-26) \frac{\eta}{r^3},
\ee
where
\be
\label{eq:a4}
a_4(\eta)=\eta\left(\frac{94}{3}-\frac{41}{32}\pi^2\right).
\ee
Note that the 1PN contribution to $A(r)$ is exactly zero. We have also included a pseudo-4PN contribution to $A(r)$, where the coefficient is parametrized as \cite{buonanno-caltechEOB09}
\be
\label{eq:a5}
a_5(\eta) = \eta (\lambda_0 + \lambda_1 \eta).
\ee
For our calculation of the EOB ISCO we shall not need to make use of the functions $B(r)$ or $D(r)$.

We further define the functions $A^{\rm T2PN}(r)$, $A^{\rm T3PN}(r)$, and $A^{\rm T4PN}(r)$ as the Taylor expansion \eqref{eq:A-taylor} truncated at 2PN, 3PN, or 4PN order. We also define the following Pad\'{e} approximants \cite{damour-jaranowski-schafer-EOBisco} to $A(r)$ which are listed in Eqs.~(50)\mbox{--}(56) of Ref.~\cite{boyle-etal-Efluxcomparison}:
\bs
\label{eq:Apade}
\be
\label{eq:A2pnpade}
A^{\rm P2PN}\equiv P_2^1[A^{\rm 2PN}] = \frac{r(-4+2r+\eta)}{2r^2 + 2\eta +r\eta} \;\; \text{at 2PN order,}
\ee
\be
\label{eq:A3pnpade}
A^{\rm P3PN}\equiv P_3^1[A^{\rm 3PN}] = \frac{{\rm Num}(A^1_3)}{{\rm Den}(A^1_3)} \;\; \text{at 3PN order, with}
\ee
\be
{\rm Num}(A^1_3)=r^2[a_4(\eta) + 8\eta -16 + r(8-2\eta)],
\ee
\begin{multline}
{\rm Den}(A^1_3)=r^3(8-2\eta) + r^2[a_4(\eta)+4\eta] \\ + r[2a_4(\eta)+8\eta] + 4[\eta^2 + a_4(\eta)], \;\; \text{and}
\end{multline}
\be
\label{eq:A4pnpade}
A^{\rm P4PN}\equiv P_4^1[A^{\rm 4PN}] = \frac{{\rm Num}(A^1_4)}{{\rm Den}(A^1_4)} \;\; \text{at 4PN order, with}
\ee
\begin{multline}
{\rm Num}(A^1_4)=r^3[32-24\eta - 4a_4(\eta) -a_5(\eta)] \\ + r^4[a_4(\eta)-16+8\eta],  \;\; \text{and}
\end{multline}
\begin{multline}
\label{eq:A14denom}
{\rm Den}(A^1_4)= -a_4^2(\eta) -8a_5(\eta) -8a_4(\eta) \eta +2 a_5(\eta) \eta
\\
- 16\eta^2 + r[ -8a_4(\eta) -4a_5(\eta) -2 a_4(\eta) \eta - 16\eta^2] \\
+ r^2[ -4a_4(\eta) - 2a_5(\eta) -16\eta] \\
+ r^3[ -2a_4(\eta)-a_5(\eta)-8\eta] + r^4[-16+a_4(\eta)+8\eta].
\end{multline}
\es

Note that the Taylor expansion in $u=M/r$ of the above Pad\'{e} approximants reduces to Eq.~\eqref{eq:A-taylor} at the appropriate PN order. However, the Pad\'{e} approximants of $A(r)$ also have the following interesting property: if one takes any of Eqs.~\eqref{eq:Apade} and computes the Taylor expansion not in $u$ but in $\eta$, one still arrives at the Taylor expansion of $A(r)$ in Eq.~\eqref{eq:A-taylor}.

The EOB ISCO is an inflection point in the radial motion given by [Eq.~(2.17) of \cite{EOB-BD2}]
\be
\frac{\partial H^{\rm real}}{\partial R}(R,P_R=0,{\mathcal J}) = 0 =\frac{\partial^2 H^{\rm real}}{\partial R^2}(R,P_R=0,{\mathcal J}),
\ee
where the total angular momentum ${\mathcal J} \equiv P_{\varphi}$ is fixed. This is equivalent to the system
\be
\frac{\partial \hat{H}^{\rm eff}}{\partial r}(r,p_r=0,p_{\varphi}) = 0 =\frac{\partial^2 \hat{H}^{\rm eff}}{\partial r^2}(r,p_r=0,p_{\varphi})
\ee
with $p_{\varphi}$ fixed, and this can be simplified to
\bs
\be
r_0^2 A_0'' (r_0^2 + j_0^2 ) - 4r_0 A_0' j_0^2 + 6A_0 j_0^2 =0,
\ee
\be
r_0 A_0' (r_0^2 + j_0^2) -2A_0 j_0^2 =0.
\ee
\es
Here $A'\equiv dA(r)/dr$, $j\equiv p_{\varphi}$, and a subscript $0$ refers to quantities evaluated at the ISCO. The angular momentum can be solved for explicitly, leaving a single equation that must be solved numerically to determine the ISCO radius:
\be
\label{eq:iscoradiuseqn}
r_0 A_0 A_0'' - 2 r_0 (A_0')^2 + 3 A_0 A'_0 =0.
\ee
The angular orbital frequency of the ISCO is then found from Eq.~\eqref{eq:EOB-dphidt} with $p_r=0$,
\be
\left.
M \Omega_0\equiv \frac{d\varphi}{dt}\right|_0 = \frac{A_0 j_0}{\eta r_0^2 \hat{H}^{\rm real}_0 \hat{H}^{\rm eff}_0}.
\ee

For reference we also note that the EOB ``horizon'' is determined by solving  $A(r)=0$, while the EOB ``light-ring'' is found from the roots of \cite{EOB-damour-nagar-finalspin}
\be
\label{eq:EOBlightring}
\frac{d}{dr}\left(\frac{A(r)}{r^2}\right)=0.
\ee
\subsubsection{\label{sec:logEOB}Logarithmic form of $A(r)$}
Building on previous works \cite{damour-spinningEOB,damour-jaranowski-schafer-spinEOB,barausse-racine-buonanno-PRD2009}, Barausse and Buonanno \cite{barausse-buonanno-spinEOB} have recently developed a new EOB Hamiltonian valid for spinning binaries. Their new Hamiltonian has several interesting and desirable properties. In particular, in the test-mass limit it reproduces the dynamics of the Mathisson-Papapetrou-Dixon equations for a spinning point-particle in the Kerr spacetime (incorporating spin-orbit interactions to all PN orders) \cite{papapetrou-spinningI-1951,papapetrou-spinningII-1951,mathisson-1931-original,*mathisson-1931-reprint,mathisson-1937-original,*mathisson-1937-reprint,dixon-extendedbodiesI-PRSocA1970,dixon-extendedbodiesII-PRSocA1970,dixon-extendedbodiesIII-PRSocA1974,barausse-racine-buonanno-PRD2009}; and its PN expansion (for any mass ratio) reproduces the 3PN point-particle Hamiltonian, as well as the leading-order PN spin-spin interaction and the spin-orbit interaction up to 2.5 PN order.

The details of this improved spinning EOB Hamiltonian are quite complicated, but in the nonspinning limit the real and effective Hamiltonians match the forms given in the previous section. However, Barausse and Buonanno \cite{barausse-buonanno-spinEOB} have employed a different form for the functions $A(r)$ and $D(r)$ that appear in the effective metric. Rather than using Pad\'{e} resummations of these functions, they introduce a logarithmic dependence which improves the behavior of $A(r)$ in the spinning case. More specifically, \cite{yi-etal-caltechEOB09,barausse-buonanno-spinEOB} found that when spins were present, the 4PN and 5PN Pad\'{e} versions of $A(r)$ contain poles. Also the Pad\'{e} resummation of $A(r)$ did not always guarantee the existence of an ISCO in the spinning case, and when it did the ISCO did not vary monotonically with the spin magnitude. In the nonspinning limit, their new form for $A(r)$ reduces to
\begin{multline}
\label{eq:A-log}
A^{\rm log}(r) = (1-\eta K)^{-2} [1-2 u (1-\eta K)] \\
\times
[ 1 + \log(1+\Delta_1 u + \Delta_2 u^2 + \Delta_3 u^3 + \Delta_4 u^4)],
\end{multline}
where $u\equiv M/r$, $\log$ refers to the natural logarithm, and the coefficients $\Delta_0$ through $\Delta_4$ are given in Eqs.~(5.77)\mbox{--}(5.81) of \cite{barausse-buonanno-spinEOB} (with $a=0$) and are functions of $\eta$ and $K$. The function $K=K(\eta)$ parametrizes 4PN (and higher-order) corrections in Eq.~\eqref{eq:A-log}. It is given by [Eq.~(6.11) of \cite{barausse-buonanno-spinEOB}]
\be
K(\eta) = K_0 (1-4\eta)^2 + 4(1-2 \eta)\eta,
\ee
where the constant $K_0=1.4467$ is chosen such that the resulting $A^{\rm log}(r)$ exactly reproduces the conservative GSF corrections to the ISCO computed by Barack and Sago \cite{barack-sago_isco}.
\subsection{\label{sec:shanks}Shanks transformation}
A final method that we will consider is the ``Shanks transformation,'' a nonlinear series acceleration method that can \emph{sometimes} increase the convergence rate of a sequence of partial sums \cite{bender-orszag}. This technique was introduced in the context of determining the ISCO in \cite{damour-jaranowski-schafer-EOBisco}. The Shanks transformation relies on the approximation that the $n$th term in a converging sequence of partial sums $Q_n$ is related to the $n\rightarrow \infty$ term $Q$ by
\be
Q_n = Q + \alpha \epsilon^n,
\ee
with $|\epsilon|<1$.
By writing out equations for three successive terms in the sequence ($Q_{n-1}$, $Q_n$, $Q_{n+1}$) one can solve for the parameters $Q$, $\alpha$, and $\epsilon$. Then for any ISCO quantity (e.g., the frequency, radius, or coefficient $c_{\Omega}^{\rm PN}$ defined below) with known values at 1PN, 2PN, and 3PN orders, we can define the Shanks transformation of that quantity by
\be
Q_{\rm isco}^{\rm S} = \frac{Q_{\rm isco}^{\rm 3PN} Q_{\rm isco}^{\rm 1PN} - (Q_{\rm isco}^{\rm 2PN})^2 }{Q_{\rm isco}^{\rm 3PN} - 2 Q_{\rm isco}^{\rm 2PN} +Q_{\rm isco}^{\rm 1PN}}.
\ee
This transformation will be applied to some of the ISCO methods discussed earlier in this section.
\section{\label{sec:results}Results}
For each of the methods reviewed in Sec.~\ref{sec:PNiscos},  I have numerically computed the dimensionless angular orbital frequency $M\Omega^{\rm PN}(\eta)$ of the ISCO as a function of $\eta$, and compared it with the renormalized Barack-Sago result [Eq.~\eqref{eq:Omega_flat_renorm}]. Specifically, I compute the analog of the coefficient $c_{\Omega}^{\rm ren} \approx 1.251$ [Eq.~\eqref{eq:Omega_flat_renorm}] via
\be
\label{eq:Omega_ren_PN}
c_{\Omega}^{\rm PN} = \lim_{\eta\rightarrow 0} \frac{1}{\eta}\left[ \frac{\Omega_{\rm isco}^{\rm PN}(\eta)}{\Omega_{\rm isco}^{\rm Schw}} -1 \right],
\ee
where the limit is taken by evaluating at some sufficiently small value of $\eta$, and $\Omega^{\rm PN}$ is different for each method. The fractional error from the exact Barack-Sago result is also computed,
\be
\label{eq:Delta-cOmega}
\Delta_{c_{\Omega}} = \frac{c_{\Omega}^{\rm PN}}{c_{\Omega}^{\rm ren}} - 1.
\ee

Several of the methods discussed do not reproduce the standard Schwarzschild test-mass ISCO; these methods are ignored when computing $c_{\Omega}^{\rm PN}$. The remaining methods for computing the ISCO are abbreviated as follows:
\begin{enumerate}
  \item $C_0{\rm 2PN}$, $C_0{\rm 3PN}$, $C_0{\rm 4PN}$: the gauge-invariant stability condition from Sec.~\ref{sec:C0invar} at 2PN and 3PN orders. The pseudo-4PN version ($C_0{\rm 4PN}$, not present in Table \ref{tab:compareisco}) fits a 4PN term to the exact BS result [see Eqs.~\eqref{eq:C0invar4PN} and \eqref{eq:c4PN} below].
  \item $E_h{\rm 1PN}$, $E_h{\rm 2PN}$, $E_h{\rm 3PN}$: the hybrid energy-function method in Sec.~\ref{sec:hybridenergy} at each PN order.
  \item KWW-1PN, KWW-2PN, KWW-3PN: the Kidder-Will-Wiseman hybrid equations-of-motion approach at each PN order (Sec.~\ref{sec:hybridpneqns}).
  \item HH-1PN, HH-2PN, HH-3PN: the hybrid-Hamiltonian method of Sec.~\ref{sec:hybridhamiltonian} at each PN order.
  \item e2PN-P, e3PN-P: the $e$-method of Sec.~\ref{sec:emethod} using the 2PN and 3PN order Pad\'{e} resummation of $e(x)$.
  \item j1PN-P, j2PN-P, j3PN-P: the $j$-method of Sec.~\ref{sec:emethod} using the Pad\'{e} resummation of $j^2(x)$ at each PN order.
  \item A2PN-T, A3PN-T: the Taylor expansion of $A(r)$ [Eq.~\eqref{eq:A-taylor}] at 2PN and 3PN orders.
  \item A4PN-${\rm T}_A$, A4PN-${\rm T}_B$: the 4PN Taylor expansion of $A(r)$, with the two choices of the pseudo-4PN coefficient $a_5$ suggested in \cite{buonanno-caltechEOB09}:
\bs
\label{eq:EOBlambdafits}
\begin{align}
\text{Choice A:} &\;\; \lambda_0=25.375, \;\;\;\;\;\; \lambda_1=0 , \;\;\; \text{and} \\
\text{Choice B:} &\;\; \lambda_0=-7.3, \;\;\;\;\;\; \lambda_1=95.6.
\end{align}
\es
    \item A2PN-P, A3PN-P, A4PN-${\rm P}_A$, A4PN-${\rm P}_B$, A4PN-${\rm P}_C$: analogous to items 7 and 8 above, but using the Pad\'{e} approximants of $A(r)$ listed in Eqs.~\eqref{eq:Apade}. A4PN-${\rm P}_C$ uses a fit for the pseudo-4PN parameter $a_5$ that exactly reproduces the BS conservative GSF ISCO shift [see Eq.\eqref{eq:lambdafit} below].
    \item AlogBB: denotes the logarithmic form of $A(r)$ \cite{barausse-buonanno-spinEOB} given in Eq.~\eqref{eq:A-log}; it is not listed in Table \ref{tab:compareisco} because it exactly reproduces the BS value by construction.
    \item HH-S, $E_h$-S, KWW-S, and j-P-S all denote the Shanks transformation applied to the hybrid-Hamiltonian, hybrid energy-function, Kidder-Will-Wiseman, and $j$-methods, using the values for $c_{\Omega}^{\rm PN}$ for these methods at 1PN, 2PN, and 3PN orders.
    \item E1PN, E2PN, E3PN: uses the standard PN circular-orbit energy in Eq.~\eqref{eq:E3PN} at 1PN, 2PN, and 3PN orders to compute the ISCO. E-S denotes the Shanks transformation applied to the PN circular-orbit energy using the values from E1PN, E2PN, and E3PN.
\end{enumerate}

\begin{figure}[t]
\includegraphics[angle=0, width=0.48\textwidth]{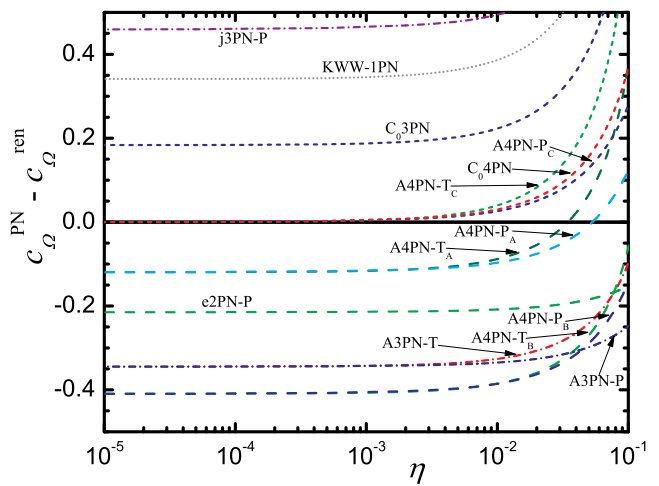}
\caption{\label{fig:dcOmega}(color online). The difference between various PN methods and the renormalized Barack-Sago conservative correction to the ISCO frequency $c_{\Omega}^{\rm ren}$. We show only the methods from Table \ref{tab:compareisco} that have reasonable agreement (within $\sim 50\%$) with the exact value for $c_{\Omega}^{\rm ren}$ [Eq.~\eqref{eq:cOmegaren}].}
\end{figure}

\begin{figure*}[t]
$
\begin{array}{cc}
\includegraphics[angle=0, width=0.48\textwidth]{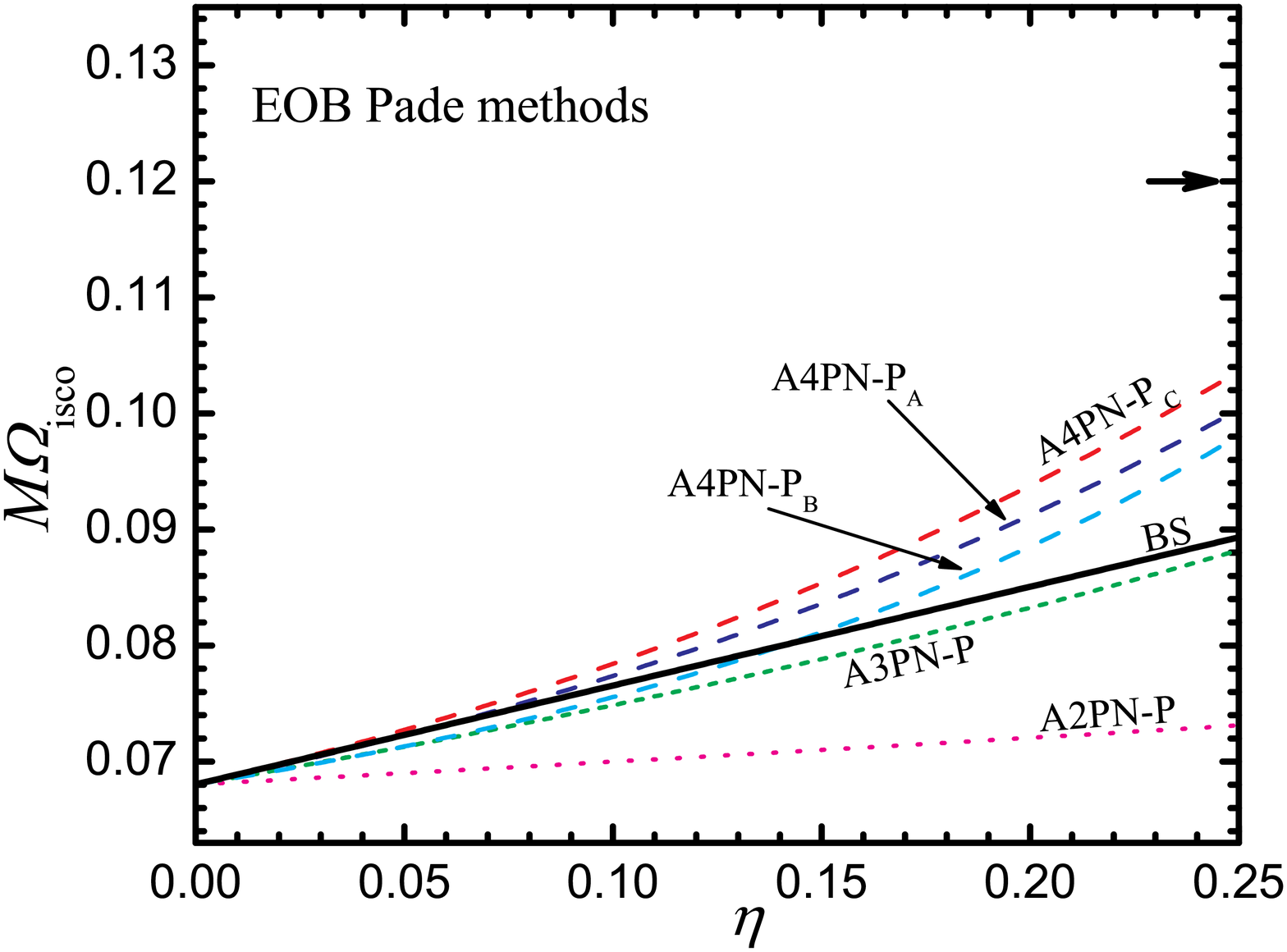} &
\includegraphics[angle=0, width=0.48\textwidth]{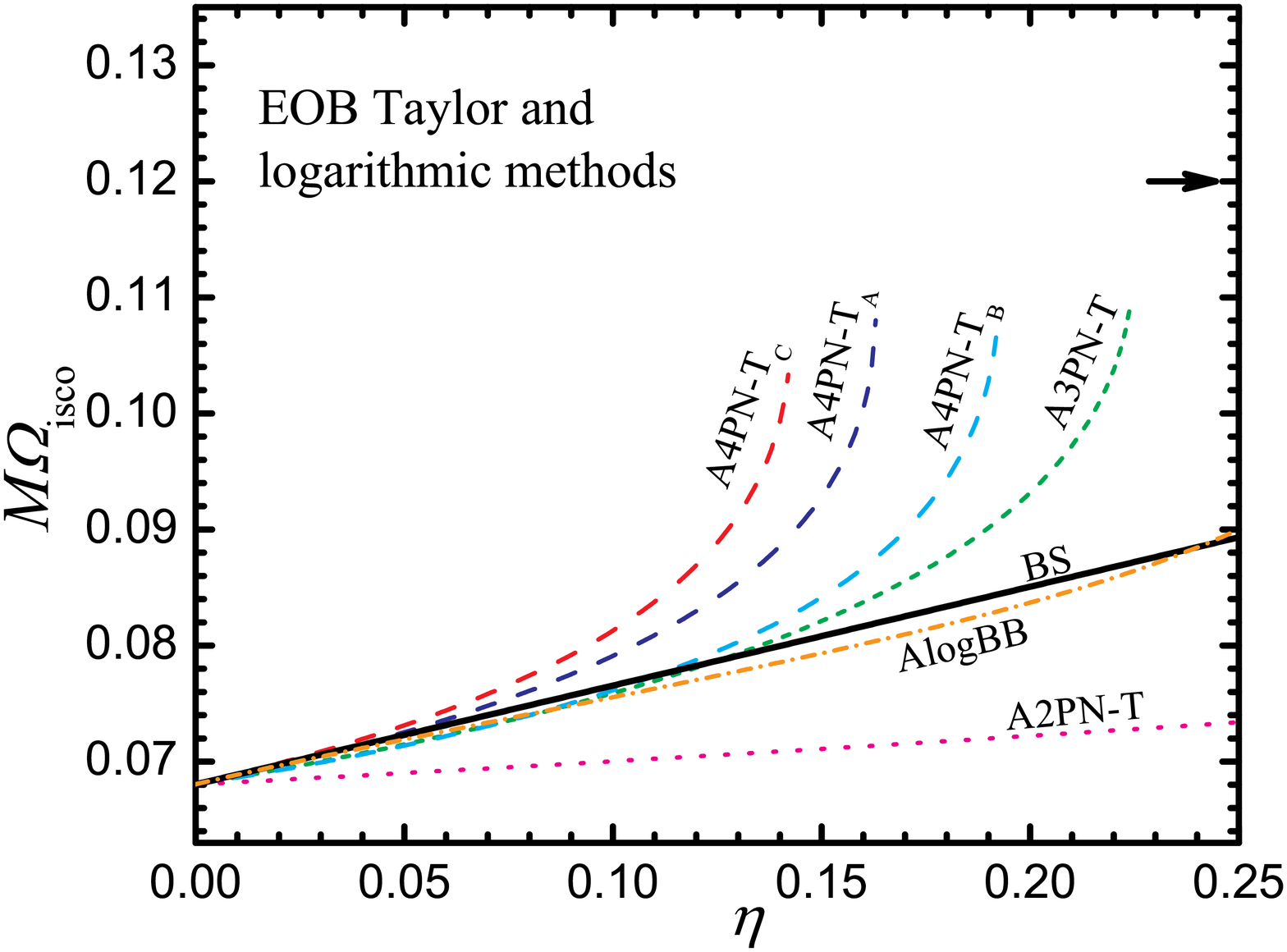} \\
\includegraphics[angle=0, width=0.48\textwidth]{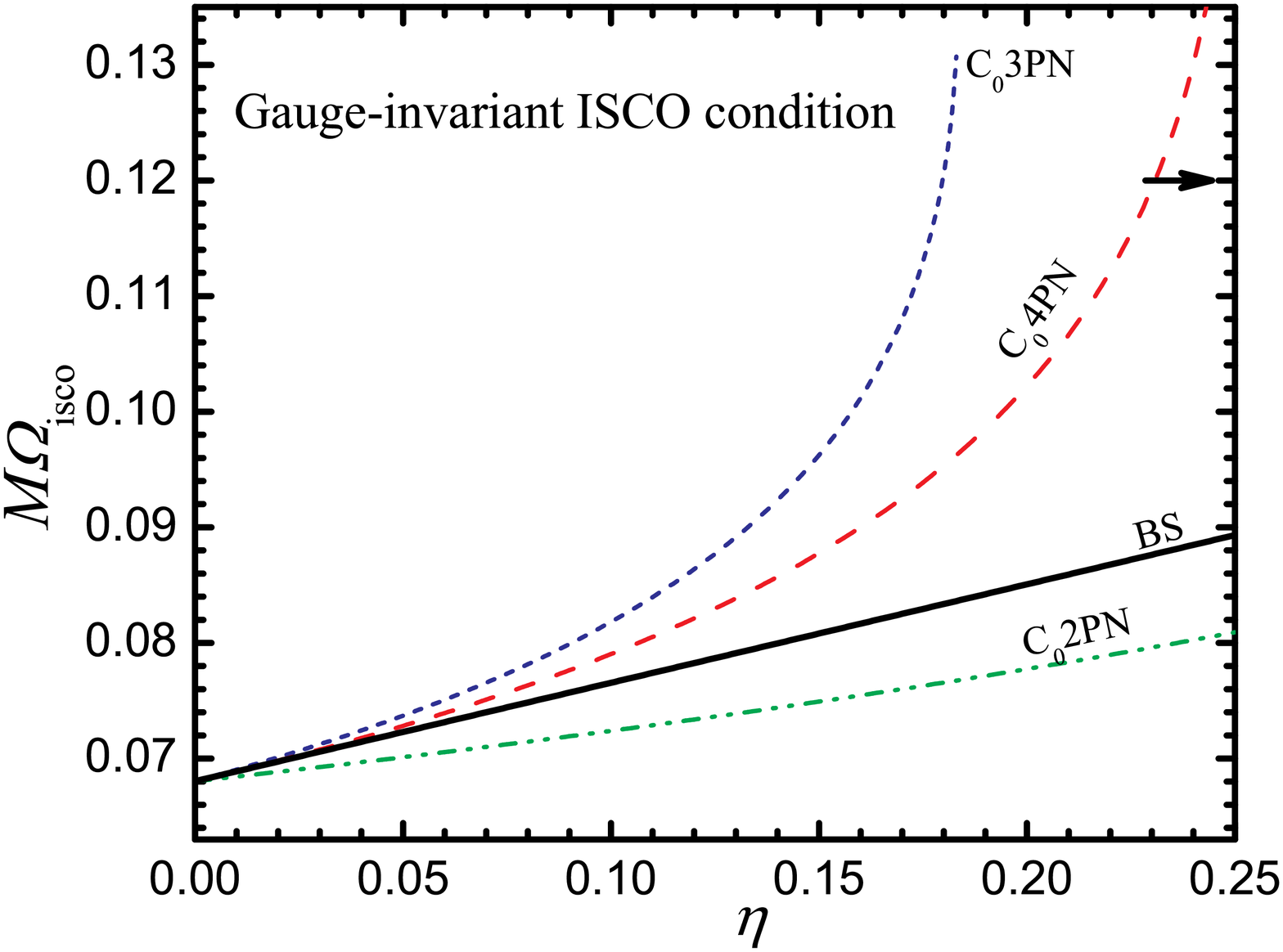} &
\includegraphics[angle=0, width=0.48\textwidth]{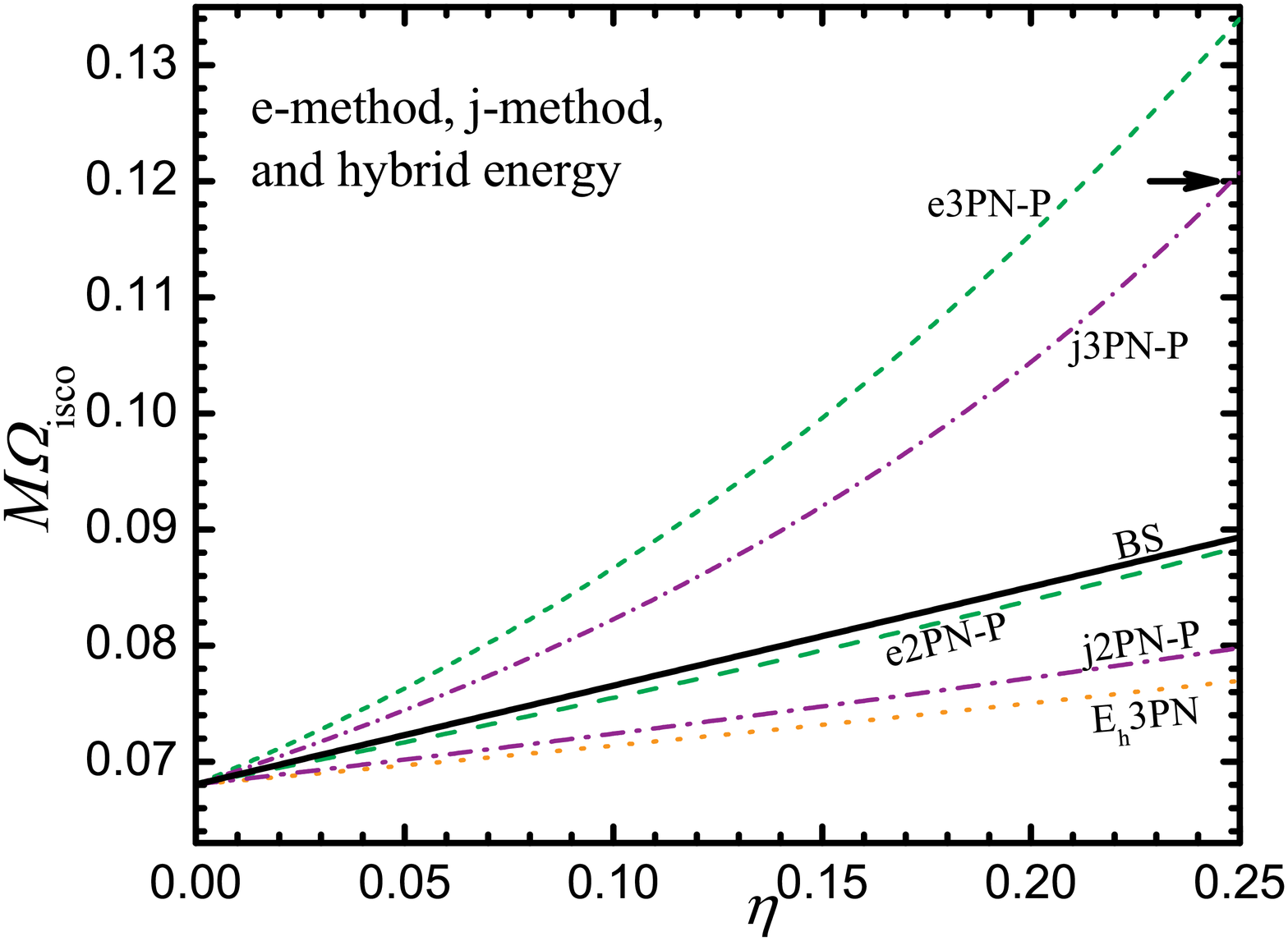}
\end{array}
$
\caption{\label{fig:iscofreq}(color online). ISCO orbital angular frequency as a function of the reduced mass ratio $\eta$. The solid black line in each plot is the renormalized Barack-Sago result (extended to large $\eta$). The abbreviations for each of the methods used are explained in the text. All EOB methods that use Pad\'{e} approximants for $A(r)$ are shown in the upper-left, and those that use Taylor expansions for $A(r)$ (as well as the logarithmic approach of \cite{barausse-buonanno-spinEOB}) are in the upper-right. The lower-left plot shows the ISCO computed from different PN orders of the gauge-invariant ISCO condition in Sec.~\ref{sec:C0invar}. The lower-right plot shows results for the $e$-method, $j$-method, and the 3PN hybrid energy function. The arrow near the point $(0.25,0.12)$ indicates the equal-mass ISCO from quasicircular initial data calculations in \cite{caudill-etal-initialdata-PRD2006}.}
\end{figure*}

The resulting values for $c_{\Omega}^{\rm PN}$ and $\Delta_{c_{\Omega}}$ are listed in Table \ref{tab:compareisco}.
Figure \ref{fig:dcOmega} illustrates how some of the better-performing PN methods deviate from the exact BS value $c_{\Omega}^{\rm ren}$ as a function of $\eta$. Figure \ref{fig:iscofreq} shows the ISCO frequency for large values of $\eta$ for several methods.

Table \ref{tab:iscoequalmass} lists the ISCO frequency in the equal-mass case for several of the methods discussed in Sec.~\ref{sec:PNiscos}, along with their fractional errors from the QCID results of \cite{caudill-etal-initialdata-PRD2006} [$\Omega_{\rm isco}^{\rm QCID}(1/4)\approx0.12$].\footnote{The values for the equal-mass ISCO in \cite{caudill-etal-initialdata-PRD2006} vary from $0.121$ to $0.124$ depending on the choice of method or boundary condition (the average is $0.122$; see Table II of \cite{caudill-etal-initialdata-PRD2006}). The comparisons in the second column of Table \ref{tab:iscoequalmass} here drop the third uncertain digit.} This value was chosen because (to my knowledge) Ref.~\cite{caudill-etal-initialdata-PRD2006} seems to be the most recent and precise study of the ISCO using QCID calculations. However it is not at all clear if this value accurately represents the ``true'' ISCO in the equal-mass case. This is especially true in light of the assumption of spatial conformal-flatness used in \cite{caudill-etal-initialdata-PRD2006}; in a PN-context the spatial-metric is known to be conformally-flat only to 1PN order. One should thus interpret the PN comparisons in Table \ref{tab:iscoequalmass} with caution and with the understanding that the exact value of the equal-mass ISCO is not accurately known (unlike the case of the BS conservative ISCO shift \cite{barack-sago_isco,barack-sago-eccentricselfforce}). Nonetheless, I believe that the ISCO value quoted above from \cite{caudill-etal-initialdata-PRD2006} represents our current best-guess, so I will use that value in the remainder of this paper.

Shortly after this article was accepted for publication, I became aware of an analysis of the ISCO using the ``skeleton'' approximation of \cite{faye-jaranowski-schafer-skeletonapprox-PRD2004}, a truncation of Einstein's equations that assumes conformal flatness and drops some gravitational-field energy terms. The resulting circular-orbit energy function that is derived from this approximation is computed to 10PN order; it agrees with the test-particle limit to all PN orders, but only agrees with the standard PN approximation to 1PN order for finite-$\eta$. The equal-mass ISCO frequency computed from this 10PN-order energy function (see Sec.~VI of \cite{faye-jaranowski-schafer-skeletonapprox-PRD2004}) is $0.0544$, differing considerably from the $0.122$ value of \cite{caudill-etal-initialdata-PRD2006}. I have also computed a hybrid version of this energy function (along the lines of Sec.~\ref{sec:hybridenergy}); the resulting value for $c_{\Omega}^{\rm PN}$ (at the 10PN level) is $\approx -2.86$, significantly different from the true value. Because these energy functions (and their lower PN-order variants) are based on a truncation of Einstein's equations and do not perform better than the top several approaches in Tables \ref{tab:compareisco} and \ref{tab:iscoequalmass}, I do not consider them further here.
\begin{table}[t]
\caption{\label{tab:compareisco}Tabulation of the conservative GSF correction to the ISCO for several of the PN methods presented in the text. The first column lists the PN method used (descriptions of the abbreviations are given in Sec.~\ref{sec:results}). The second column lists the coefficient $c_{\Omega}^{\rm PN}$ defined in Eq.~\eqref{eq:Omega_ren_PN}. This was computed for $\eta=10^{-6}$; for smaller $\eta$ the values are unchanged to four significant figures. These values are compared with the exact (renormalized) Barack-Sago result, $c_{\Omega}^{\rm ren}=1.251$, in the third column, which lists the fractional error [Eq.~\eqref{eq:Delta-cOmega}]. The table is sorted by the absolute value of the fractional error in the third column (most accurate method first).}
\begin{ruledtabular}
\begin{tabular}{ldd}
Method & c_{\Omega}^{\rm PN}  & \Delta_{c_{\Omega}} \\
\hline
A4PN-${\rm P}_A$ & 1.132 & -0.0955 \\
A4PN-${\rm T}_A$ & 1.132 & -0.0955 \\
$C_0{\rm 3PN}$ & 1.435 & 0.1467 \\
e2PN-P & 1.036 & -0.1717 \\
KWW-1PN & 1.592 & 0.2726 \\
A3PN-P & 0.9067 & -0.2754 \\
A3PN-T & 0.9067 & -0.2754 \\
A4PN-${\rm P}_B$ & 0.8419 & -0.3272\\
A4PN-${\rm T}_B$ & 0.8419 & -0.3272 \\
j3PN-P & 1.711 & 0.3671 \\
j2PN-P & 0.6146 & -0.5088 \\
KWW-S & 0.5610 & -0.5515 \\
$C_0{\rm 2PN}$ & 0.5833 & -0.5338 \\
$E_h{\rm 3PN}$ & 0.4705 & -0.6240 \\
e3PN-P & 2.178 & 0.7409 \\
A2PN-P & 0.2794 & -0.7767 \\
A2PN-T & 0.2794 & -0.7767 \\
$E_h{\rm 2PN}$ & 0.0902 & -0.9279 \\
$E_h{\rm 1PN}$ & -0.014\,73 & -1.011 \\
$E_h$-S & -0.054\,71 & -1.044 \\
HH-S & -0.1486 & -1.119 \\
j1PN-P & -0.1667 & -1.133 \\
KWW-2PN & -1.542 & -2.232 \\
j-P-S & -2.104 & -2.682 \\
KWW-3PN & 4.851 & 2.877 \\
HH-1PN & 6.062 & 3.844 \\
HH-2PN & -12.75 & -11.19 \\
HH-3PN & 25.42 & 19.32 \\
\end{tabular}
\end{ruledtabular}
\end{table}
\begin{table}[t]
\caption{\label{tab:iscoequalmass}ISCO frequency for equal-mass binaries for selected methods presented in the text. The first column lists the PN method used (descriptions of the abbreviations are given in Sec.~\ref{sec:results}). The second column lists the ISCO angular orbital frequency $M\Omega_{\rm isco}(\eta=1/4)$. The third column lists the fractional error from the approximate QCID value $\Omega_{\rm isco}^{\rm QCID} \approx 0.12$ reported in \cite{caudill-etal-initialdata-PRD2006}. The table is sorted by the absolute value of the fractional error listed in the third column.}
\begin{ruledtabular}
\begin{tabular}{ldd}
Method & \Omega_{\rm isco}^{\rm PN}  & \Delta_{\Omega}^{\rm QCID} \\
\hline
j3PN-P & 0.1207 & 0.0061 \\
E-S & 0.1285 & 0.071 \\
E3PN & 0.1287 & 0.073 \\
e3PN-P & 0.1340 & 0.12 \\
A4PN-${\rm P}_C$ & 0.1036 & -0.14 \\
E2PN & 0.1371 & 0.14 \\
A4PN-${\rm P}_A$ & 0.1004 & -0.16 \\
A4PN-${\rm P}_B$ & 0.098\,07 & -0.18 \\
AlogBB & 0.089\,99 & -0.25 \\
e2PN-P & 0.088\,50 & -0.26 \\
A3PN-P & 0.088\,22 & -0.26 \\
$C_0$4PN & 0.1567 & 0.31 \\
$C_0$2PN & 0.0809 & -0.33 \\
j2PN-P & 0.079\,80 & -0.33 \\
$E_h$3PN & 0.076\,98 & -0.36 \\
A2PN-T & 0.073\,40 & -0.39 \\
A2PN-P & 0.073\,12 & -0.39 \\
$E_h$2PN & 0.069\,59 & -0.42 \\
$E_h$1PN & 0.067\,79 & -0.44 \\
$E_h$-S & 0.067\,21 & -0.44 \\
j1PN-P & 0.065\,30 & -0.46 \\
j-P-S & 0.057\,35 & -0.52 \\
E1PN & 0.5224 & 3.4 \\
\end{tabular}
\end{ruledtabular}
\end{table}
\section{\label{sec:discuss}Discussion}
From the values listed in Tables \ref{tab:compareisco} and \ref{tab:iscoequalmass} we now make the following observations:
\begin{enumerate}
  \item Nearly all possible methods for computing finite-mass ratio corrections to the ISCO in the PN framework were considered, and these methods generally fall into two categories: nonresummed and resummed approaches. For the purpose of determining the conservative ISCO shift for very small $\eta$, all but one of the nonresummed approaches is useless for computing $c_{\Omega}^{\rm PN}$ since they generally do not reproduce the exact Schwarzschild ISCO.

  \item All of the methods discussed have appeared previously in the literature, although not all were previously investigated at 3PN order. In particular, note that the Kidder-Will-Wiseman \cite{kidderwillwiseman-transition-PRD1993} hybrid approach (which originally motivated the development of resummation methods) was here extended to 3PN order. In contrast to the 2PN order results reported in \cite{kidderwillwiseman-transition-PRD1993}, at 3PN order the conservative ISCO shift at least has the correct sign. However, the fact that the 1PN version (KWW-1PN) makes the most accurate KWW prediction for $c_{\Omega}^{\rm PN}$ and the 3PN version (KWW-3PN) the least accurate, further suggests \cite{wex-schafer-CQG1993,damour-iyer-sathyaprakash-PRD1998} that the KWW hybrid approach is not a well-behaved resummation method. This pattern also occurs for the hybrid-Hamiltonian (HH) method (Sec.~\ref{sec:hybridhamiltonian}) and the $e$-method (Sec.~\ref{sec:emethod}), suggesting that they too are not preferred approaches. This is in contrast with the remaining methods listed in Tables \ref{tab:compareisco} and \ref{tab:iscoequalmass}, which share the property that the higher PN iteration of a given method produces a value closer to the exact result.

  \item The method that produces the best agreement with the exact result ($\sim 10\%$ error) is the EOB method in which a pseudo-4PN parameter $a_5(\eta)$ is introduced and its value is adjusted to NR simulations in \cite{buonanno-caltechEOB09}. In particular, only one of the two suggested choices \cite{buonanno-caltechEOB09} for $a_5$ [choice~A in Eq.~\eqref{eq:EOBlambdafits}] gives good agreement. Choice~B gives an error that is 3 times worse. It is especially interesting that the fits to the NR simulations---which are done in the $q\sim 1$ limit---have in some sense ``preselected'' a value for $a_5$ that is closest to reproducing the exact result of a $q \ll 1$ calculation.

  \item It is interesting to note that if we neglect the methods that involve some sort of fitting to numerical results, then the best EOB approach (A3PN-P) is not the most accurate method. Rather, in both the extreme-mass ratio (Table \ref{tab:compareisco}) and equal-mass (Table \ref{tab:iscoequalmass}) cases, two distinct nonresummed approaches based on the ordinary 3PN equations of motion are among the most accurate approaches.

  \item However, note also that in both the extreme-mass ratio and equal-mass cases, the error associated with the A3PN-P method is nearly the same ($\sim 27\%$), and arises from a single, distinct method. Introducing a pseudo-4PN term and calibrating to NR (A4PN-${\rm P}_A$) or to the BS results (A4PN-${\rm P}_C$) further reduces the errors in both mass-ratio limits.

   \item In the equal-mass case (where the nonresummed PN series has good convergence properties), the ISCO computed from the 3PN circular-orbit energy shows remarkable agreement ($\sim 7\%$) with the QCID result from \cite{caudill-etal-initialdata-PRD2006}---better than any EOB method.

   \item The $j$-method at 3PN order (j3PN-P) produces nearly exact agreement with the equal-mass QCID ISCO. This is possibly coincidental, and partly due to the truncation of the ``exact'' QCID result to two digits.
       However, we also note that j3PN-P does moderately well at reproducing the BS ISCO shift ($\sim 37\%$ error), and the different PN iterations of the $j$-method (1PN, 2PN, 3PN) show successive improvement at each PN order in both the equal-mass and extreme-mass-ratio cases (in contrast with the $e$-method; see Tables \ref{tab:compareisco} and \ref{tab:iscoequalmass}).

  \item {\label{itemC0}} Surprisingly, the method which best reproduces the Barack-Sago conservative GSF ISCO corrections (without relying on any fits with NR or GSF calculations) is the gauge-invariant ISCO condition $C_0$3PN of \cite{blanchetiyer3PN}. This is the only nonresummed approach in Table \ref{tab:compareisco}. It is an especially interesting method because it both reproduces the exact test-mass ISCO \emph{and} matches the conservative GSF ISCO corrections with good accuracy ($\sim 15\%$)---nearly twice the accuracy of the best 3PN EOB approach. On the other hand, the $C_0$3PN method does not produce an ISCO for $\eta>0.183$, while the A3PN-P method yields a well-defined ISCO for any $\eta\leq 1/4$.

  \item Notice also that in Table \ref{tab:compareisco}, the Taylor and Pad\'{e} forms of the EOB potential give identical results at a given PN order. This arises from the fact [discussed after Eq.~\eqref{eq:A14denom}] that the $\eta$-expansion of the Pad\'{e} approximants of $A(r)$ reduces to the Taylor expansion given in Eq.~\eqref{eq:A-taylor}.

  \item Regarding the logarithmic form of $A(r)$ proposed in \cite{barausse-buonanno-spinEOB}, note that the AlogBB ISCO frequency maintains a nearly constant slope for all $\eta$, closely following the large-$\eta$ extrapolation of the BS result to which it is calibrated (Fig.~\ref{fig:iscofreq}).

  \item Regarding the Shanks transformation: while it was only possible to apply it for methods with ISCO quantities defined at 1PN, 2PN, and 3PN orders, for the methods investigated in Tables \ref{tab:compareisco} and \ref{tab:iscoequalmass} it generally did not yield an improvement in accuracy. The exception is for the standard PN-energy function method E3PN, which saw a very slight improvement in the accuracy of the equal-mass ISCO.

  \item Also note the much larger spread in the error values listed in Table \ref{tab:compareisco} versus those in Table \ref{tab:iscoequalmass}. This is likely a reflection of the well-known poor convergence of the PN series in the small-$\eta$ limit and the relatively good convergence in the equal-mass limit.
\end{enumerate}

Let us now elaborate on point \ref{itemC0} above. It is rather curious that the gauge-invariant ISCO condition of Blanchet and Iyer \cite{blanchetiyer3PN} [Eq.~\eqref{eq:C0invar}] not only exactly reproduces the test-mass ISCO, $x_{\rm isco}=1/6$, but also provides very close agreement with the BS result for $c_{\Omega}^{\rm ren}$. Damour \cite{damour-GSF} also finds a value for $c_{\Omega}^{\rm PN}$ equal to that produced by $C_0$3PN; but his result is derived by performing a PN expansion of quantities that appear in the EOB effective metric, so the agreement with the Schwarzschild ISCO in his approach is by construction [see his Eq.~(5.41) and the associated discussion]. In Blanchet and Iyer's \cite{blanchetiyer3PN} derivation, the agreement with the test-mass ISCO was not enforced (and is thus surprising since standard PN calculations typically do not exactly reproduce strong-field results).

As for the fact that the value $c_{\Omega}^{\rm C_03PN}\approx 1.435$ [from $C_0$3PN or Damour's \cite{damour-GSF} Eq.~(5.41)] agrees more closely with the BS result than the A3PN-P value, Damour \cite{damour-GSF} suggests that this is an accident arising from the failure of certain terms in the PN expansion of EOB quantities to cancel with higher-order terms. However, it is not clear how (or if) this argument translates to an ``explanation'' of the value of  $c_{\Omega}^{\rm C_03PN}\approx 1.435$ when it is derived via the standard PN approach in Blanchet and Iyer \cite{blanchetiyer3PN}.

To further explore the possibility that the behavior of the gauge-invariant ISCO condition $C_0$3PN might be accidental, I have extended the calculation of Blanchet and Iyer \cite{blanchetiyer3PN} to the case of nonprecessing, spinning binaries. The details are presented in \cite{favata-PNspinisco}. The derivation follows that in \cite{kidderwillwiseman-transition-PRD1993,blanchetiyer3PN}, except that I also include the spin-orbit terms at 1.5PN and 2.5PN orders \cite{faye-buonanno-luc-higherorderspinI,faye-buonanno-luc-higherorderspinII}, and the spin-spin and quadrupole-monopole terms at 2PN order \cite{racine-buonanno-kidder-spinningrecoil}. The result is a condition for the ISCO which generalizes Eq.~\eqref{eq:C0invar},
\begin{multline}
\label{eq:C0hat-spin-mod}
\hat{C}_0 \equiv \frac{M^2}{x_0^3} C_0 = 1- 6x_0 \\
+ x_0^{3/2} \left( 14 \frac{S^{\rm c}_{\ell}}{M^2} + 6 \frac{\delta m}{M} \frac{\Sigma^{\rm c}_{\ell}}{M^2} \right) + x_0^2 \left[ 14\eta - 3 \left( \frac{S^{\rm c}_{0,\ell}}{M^2} \right)^2 \right]
\\
+ x_0^{5/2} \left[ - \frac{S^{\rm c}_{\ell}}{M^2} (22+32\eta) - \frac{\delta m}{M} \frac{\Sigma^{\rm c}_{\ell}}{M^2} (18+15\eta) \right]
\\
+ x_0^3 \left[ \left( \frac{397}{2} - \frac{123}{16}\pi^2  \right) \eta - 14 \eta^2 \right],
\end{multline}
where $S^{\rm c}_{\ell}={\bm \ell} \cdot ({\bm S}^{\rm c}_1 + {\bm S}^{\rm c}_2)$, $\Sigma^{\rm c}_{\ell}=M {\bm \ell} \cdot ({\bm S}^{\rm c}_2/m_2 - {\bm S}^{\rm c}_1/m_1)$, $S^{\rm c}_{0,\ell}=M {\bm \ell}\cdot [(1+m_2/m_1){\bm S}^{\rm c}_1 +(1+m_1/m_2) {\bm S}^{\rm c}_2]$, ${\bm \ell}$ is the unit vector in the direction of the Newtonian orbital angular momentum, and $\delta m = m_1-m_2$.\footnote{The spin angular-momentum vectors ${\bm S}_i^{\rm c}$ used above refer to the constant-magnitude spin variables \cite{faye-buonanno-luc-higherorderspinII,kidder-spineffects,racine-buonanno-kidder-spinningrecoil}. However, the same test-mass limit [Eq.~\eqref{eq:C0spin2}] is found if the spin variables with varying magnitudes \cite{faye-buonanno-luc-higherorderspinI,faye-buonanno-luc-higherorderspinII} are used (see \cite{favata-PNspinisco}).}

If the larger BH has spin $|{\bm S}^{\rm c}_2|=\chi_2^{\rm c} m_2^2$ (and ${\bm S}^{\rm c}_1=0)$, the above condition reduces, in the test-particle limit, to
\be
\label{eq:C0spin2}
\hat{C}_0 = 1 - 6 x_0 + 8 \chi_2^{\rm c} x_0^{3/2} - 3 (\chi_2^{\rm c})^2 x_0^2 - 4 \chi_2^{\rm c} x_0^{5/2} + O(x_0^3).
\ee
A similar criterion for the exact (asymptotic) Kerr ISCO frequency parameter $X_0\equiv (m_2 \Omega^{\rm kerr}_{\rm isco})^{2/3}$ can be derived from the minimum of the orbital energy of a particle in the Kerr spacetime \cite{bptkerr}. This criterion, when expanded for small BH spin, takes the form \cite{favata-PNspinisco}
\begin{multline}
\label{eq:C0kerr-smallspin}
\hat{C}_0^{\rm kerr} = 1 - 6 X_0 + \chi_2 ( 8 X_0^{3/2} - 4X_0^{5/2} )
\\
+ \chi_2^2 \left(- 3X_0^2 + 8X_0^3 -10 X_0^4/3 \right) + O(\chi_2^3).
\end{multline}
Note that no PN expansion was used to derive this expression; it is exact to $O(\chi_2^2)$ in the BH spin.

Except for the 3PN and 4PN order spin terms (whose forms in the PN equations of motion are not currently known), Eqs.~\eqref{eq:C0spin2} and \eqref{eq:C0kerr-smallspin} agree exactly when we identify the Kerr spin parameter $\chi_2$ with $\chi_2^{\rm c}$ (note that $x_0 \rightarrow X_0$ in the test-mass limit). The fact that the extension of the gauge-invariant ISCO condition $C_0$3PN \cite{blanchetiyer3PN} to spinning BHs reproduces the exact Kerr ISCO (to the expected order in $\chi_2$) suggests (but does not prove) that the close agreement of $C_0$3PN with the conservative GSF ISCO shift is \emph{not} accidental. It remains to be seen if the predictions of Eq.~\eqref{eq:C0hat-spin-mod} will also produce good agreement with the conservative GSF ISCO shift when it is eventually calculated for Kerr BHs. In addition to this good agreement with the Kerr ISCO, condition \eqref{eq:C0hat-spin-mod} can also be shown to exactly reproduce the fully-relativistic conservative shift in the ISCO due to the spin of a small test-mass. These issues are discussed further in \cite{favata-PNspinisco}.
\section{\label{sec:NR}Joint constraints between numerical relativity, post-Newtonian theory, and self-force calculations}
The primary motivation for GSF calculations is to produce accurate waveform models for EMRIs. However, as we explore below,  GSF calculations can also improve our understanding of comparable-mass ($1/10 \lesssim q\lesssim 1$) waveforms. For example, GSF calculations can help calibrate EOB waveforms (Sec.~\ref{sec:fit4pn}; see also \cite{damour-GSF,barausse-buonanno-spinEOB,barack-damour-sago_periastron}), phenomenological inspiral-merger-ringdown (IMR) waveforms (Sec.~\ref{sec:phenomen}), and high-order terms in PN quantities like the orbital energy (Sec.~\ref{sec:fitEpn}).
In Sec.~\ref{sec:QCID} I also discuss how numerical computations of sequences of quasicircular initial data, combined with full NR calculations of the energy flux for a few orbits, could be used to estimate high-PN-order terms in the inspiral phasing.
\subsection{\label{sec:fit4pn}Fitting pseudo-4PN parameters}
In addition to the two choices for $a_5$ in Eq.~\eqref{eq:EOBlambdafits}, we can also attempt to \emph{fix} the value of $a_5$ such that the Barack-Sago conservative GSF ISCO shift is exactly reproduced. To do this we assume\footnote{We also ignore the possibility of any log terms that might be present at 4PN order; see \cite{damour-GSF,barack-damour-sago_periastron} for further discussion.} that $\lambda_1=0$ (since the GSF calculations are only to leading-order in $\eta$), and adjust $\lambda_0$ (holding $\eta$ fixed at $10^{-7}$) until the resulting  $c_{\Omega}^{\rm PN}$ matches the exact result. This is done for both the 4PN Taylor and 4PN Pad\'{e} expansions of $A(r)$ (denoted A4PN-${\rm T}_C$ and A4PN-${\rm P}_C$ here and in Figs.~\ref{fig:dcOmega} and \ref{fig:iscofreq}). The resulting value for $\lambda_0$ is
\be
\label{eq:lambdafit}
\lambda_0^{\rm fit} = 38.84
\ee
for both A4PN-${\rm T}_C$ and A4PN-${\rm P}_C$.\footnote{One can also attempt to leave undetermined the 3PN coefficient $a_4$ [Eq.~\eqref{eq:a4}] and ``fit'' it to $c_{\Omega}^{\rm ren}$. The result is $a_4 \approx 28.96 \eta$, an $\approx 55\%$ error with the exact result.} Note that this value is not wildly different from $\lambda_0=25.375$, one of the values found by fitting to the NR simulations in Ref.~\cite{buonanno-caltechEOB09} [see Eq.~\eqref{eq:EOBlambdafits}]. Damour \cite{damour-GSF} finds an equivalent constraint [see his Eq.~(4.42)], although he considers a 5PN extension of the EOB potential $A(r)$ that supplements the 3PN Taylor expansion by $\delta A = \eta (a_5^D u^5 + a_6^D u^6)$ \cite{damour-nagar-PRD09}, where $u\equiv M/r$ and $a_5^D$ and $a_6^D$ are constants.

The above fit for the pseudo-4PN parameter $a_5=\lambda_0^{\rm fit} \eta$ does not necessarily contain any new physical information about the 4PN dynamics; instead it essentially ``resums'' all of the higher-order PN corrections that contribute to the conservative GSF ISCO shift and groups them into a single 4PN term.
This value for $a_5$ would not exactly coincide with some future calculation of the 4PN expansion of $A(r)$; however, if the 5PN and higher contributions to $A(r)$ are small, it is possible that the value calculated above might not be too far from the true 4PN result.

To test this notion, compare the $O(\eta)$ terms in the 4PN Taylor expansion of $A(r)$ [Eq.~\eqref{eq:A-taylor}]: $2\eta/r^3$, $a_4/r^4$, and $a_5/r^5$. Numerically evaluating these terms at $r=6$ [and using Eq.~\eqref{eq:lambdafit} for $a_5$], we get $0.0093\eta$, $0.014\eta$, and $0.005\eta$ (respectively). The fact that the 4PN term is smallest even near the ISCO suggests (but does not prove) that the 5PN term might be small in comparison to the 4PN one. See Damour \cite{damour-GSF} for an alternative argument.

Regardless of its agreement with respect to the true value, the above choice [Eq.~\eqref{eq:lambdafit}] for the pseudo-4PN term in $A(r)$ can be said to accurately reproduce an important strong-field feature of the conservative dynamics. It might therefore be useful to fix $\lambda_0$ to the value $38.84$ in future EOB/NR comparison studies. Fixing the pseudo-4PN term in this way could also be useful for studies that use the EOB formalism to model waveforms from extreme- and intermediate-mass-ratio inspirals \cite{yunes-etal-EMRI-EOB}.

If one also had a highly accurate determination of the ISCO frequency in the equal-mass case (or any moderate mass-ratio for that matter), one could also attempt to constrain higher-order parameters in the $a_5(\eta)$ coefficient [Eq.~\eqref{eq:a5}]. For example, Ref.~\cite{caudill-etal-initialdata-PRD2006} gives values for the nonspinning, equal-mass ISCO in the range $M\Omega(1/4)= 0.121$ to $0.124$ depending on the method and boundary condition used. If we fix the value of $\lambda_0$ to that found in Eq.~\eqref{eq:lambdafit}, then we find that the above range for the ISCO implies that $\lambda_1$ must lie in the range
\be
\lambda_1 \in [488.9,665.5].
\ee
For $M\Omega(1/4)=0.122$ we find $\lambda_1=541.3$.

In addition to using the BS conservative ISCO shift, Damour \cite{damour-GSF} also used comparisons with full NR simulations to provide constraints on the parameters $a_5^D$ and $a_6^D$ discussed above. Rather than using full NR simulations (which include the dissipative dynamics), sequences of BH quasicircular initial data (combined with the conservative GSF ISCO correction) could also be used to constrain these parameters [as was done with $(\lambda_0, \lambda_1)$ above]. This has the advantage of using strictly ``conservative'' NR calculations to constrain the conservative dynamics encapsulated in $A(r)$.

We can similarly try adding a pseudo-4PN term to the gauge-invariant Blanchet-Iyer \cite{blanchetiyer3PN} stability condition discussed in Sec.~\ref{sec:C0invar},
\begin{multline}
\label{eq:C0invar4PN}
\hat{C}_0 = 1-6x_0 + 14\eta x_0^2
\\
+ \left( \left[\frac{397}{2} -\frac{123}{16}\pi^2\right]\eta -14\eta^2\right)x_0^3 + c_{\rm 4PN} \eta x_0^4,
\end{multline}
where we again ignore possible logarithmic corrections (computable from the results of \cite{blanchet-etal-selfforceII}) and assume that the 4PN terms do not modify the stability condition at $O(\eta^0)$ (allowing the exact Schwarzschild test-mass ISCO to be reproduced). Precise agreement with the Barack-Sago result is enforced by tuning $c_{\rm 4PN}$ to the value\footnote{If we perform a similar procedure leaving the $O(\eta)$ 3PN coefficient in Eq.~\eqref{eq:C0invar} undetermined, we find a value $c_{\rm 3PN} \approx 96.2$, which is within $22\%$ of the exact result.}
\be
\label{eq:c4PN}
c_{\rm 4PN} \approx -158.64.
\ee
The resulting ISCO frequency is denoted $C_0$4PN in Table~\ref{tab:iscoequalmass} and Figs.~\ref{fig:dcOmega} and \ref{fig:iscofreq}.
Evaluating the $O(\eta)$ terms in Eq.~\eqref{eq:C0invar4PN} at $x_0 \approx 1/6$,
\bs
\begin{align}
 14\eta x_0^2 &\approx 0.39\eta, \\
 \left[\frac{397}{2} -\frac{123}{16}\pi^2\right]\eta x_0^3 &\approx 0.57\eta, \\
 c_{\rm 4PN} \eta x_0^4 &\approx -0.12\eta,
\end{align}
\es
we see a similar sequence as in the EOB case with the pseudo-4PN term being smallest.
\subsection{\label{sec:phenomen}Constraints on phenomenological inspiral-merger-ringdown templates}
In the phenomenological inspiral-merger-ringdown template family developed by Ajith et~al.~\cite{ajith-etal-2009}, a frequency-domain template $\tilde{h}(f)\equiv A(f) e^{-i\Psi(f)}$ is defined in terms of a phase $\Psi(f)$ and an amplitude $A(f)$ [see Eq.~(1) of \cite{ajith-etal-2009}]. The amplitude function $A(f)$ is written as a piecewise function that transitions from an ``inspiral'' to ``merger'' at a frequency $f_1$, and to a ``ringdown'' at $f_2$. The frequency $f_1\equiv\Omega_1/\pi$  between the inspiral and merger phase is defined such that in the $\eta\rightarrow 0$ limit it reduces to the test-mass ISCO. When $\eta$ is nonzero and for initially nonspinning BHs, this transition frequency is given by [see Eq.~(2) and Table I of \cite{ajith-etal-2009}]
\be
\label{eq:Omega-phenom}
M \Omega_1 = 6^{-3/2} + \eta y^{(10)} + \eta^2 y^{(20)} + \eta^3 y^{(30)},
\ee
where $y^{(10)}=0.6437$, $y^{(20)}=-0.058\,22$, and $y^{(30)}=-7.092$. Note that $6^{3/2} y^{(10)} \approx 9.46$ and differs from $c_{\Omega}^{\rm ren}$ by a factor of about $7.6$ (although it at least has the correct sign). This lack of agreement is not surprising since the smallest mass ratio considered in \cite{ajith-etal-2009} was $0.25$ ($\eta=0.16$); their fits could only be expected to work for mass ratios larger than this. Future phenomenological IMR templates could consider fixing the value of $y^{(10)}$ to $6^{-3/2} c_{\Omega}^{\rm ren} \approx 0.085\,14$; this might help to provide better template matches with NR simulations at small mass ratios. In principle the other higher-order $\eta$ terms in Eq.~\eqref{eq:Omega-phenom} could also be fixed via any of the PN ISCO methods discussed here (see, e.g., Fig.~\ref{fig:iscofreq}) or by QCID ISCO calculations. However, it is not clear if doing so will necessarily produce a template family that can better match NR waveforms.
\subsection{\label{sec:fitEpn}Constraints on the 4PN and 5PN circular-orbit energy}
The PN energy for circular orbits [Eq.~\eqref{eq:E3PN}] contains newly computed terms \cite{blanchet-etal-selfforceII} at 4PN and 5PN orders. However, only the test-mass-limit and logarithmic pieces of these terms are known. The remaining uncertainty is parametrized by the polynomials
\bs
\label{eq:e4e5-expand}
\begin{align}
\label{eq:e4-expand}
e_4(\eta) = e_4^{(0)} + e_4^{(1)} \eta + e_4^{(2)} \eta^2 + \cdots e_4^{(p)} \eta^p, \\
\label{eq:e5-expand}
e_5(\eta) = e_5^{(0)} + e_5^{(1)} \eta + e_5^{(2)} \eta^2 + \cdots e_5^{(q)} \eta^q,
\end{align}
\es
where $p$ and $q$ are integers (probably equal to 4 and 5). Computing these functions will require the completion of the PN iteration scheme at the 4PN and 5PN levels---a daunting task. Here I attempt to partly constrain these polynomials by using numeric calculations of the ISCO from the GSF approach or quasicircular initial data calculations.

Numerical relativity calculations of the ISCO frequency for equal-mass binaries come from examining QCID sequences. The most recent results \cite{caudill-etal-initialdata-PRD2006} indicate $M\Omega_{\rm isco}(1/4)\approx 0.122$. This single value provides a possible crude constraint on PN parameters appearing in the circular orbit energy function through the ISCO condition $d/dx [\tilde{E}^{\rm PN}{\bm (}\Omega=\Omega_{\rm isco}(1/4),\eta=1/4{\bm )}]=0$, where $\tilde{E}^{\rm PN}\equiv E^{\rm PN}/(\eta M)$ is given by Eq.~\eqref{eq:E3PN}.\footnote{Keep in mind that it is not clear to what degree the QCID results represent the ``true'' ISCO embodied by the purely conservative dynamics of the full Einstein equations. This is especially true in light of the conformal-flatness assumption that is used in \cite{caudill-etal-initialdata-PRD2006}.} We can test this approach by assuming that the 3PN coefficient in $E^{\rm PN}$ is unknown and has the form
\be
-\frac{675}{64} + \eta e_3(\eta),
\ee
where the {\em known} value for $e_3(\eta)$ is
\be
e_3(\eta) = \left( \frac{34\,445}{576} -\frac{205}{96}\pi^2 \right)
- \frac{155}{96}\eta - \frac{35}{5184}\eta^2.
\ee
Ignoring the 4PN and higher order terms and using the above ISCO criterion with the equal-mass value from \cite{caudill-etal-initialdata-PRD2006} yields the constraint
\be
e_3(1/4)\approx 34.4,
\ee
which agrees with the exact value of $38.3$ to $10\%$. If we apply this procedure to the 4PN and 5PN corrections, we get the constraints
\bs
\begin{align}
\label{eq:e4cond1}
e_4(1/4) &\approx 147.3 \;\; \text{at 4PN order, and} \\
e_4(1/4) &+ 0.2952 e_5(1/4) \approx 189.6 \;\; \text{at 5PN order.}
\end{align}
\es

The reliability of these constraints is not completely clear; it depends on the accuracy and precision of the numerical simulation (including the systematic effects alluded to above), as well as on the contributions of all the higher-order PN terms. For example, suppose that the ISCO computed from \cite{caudill-etal-initialdata-PRD2006} has a fractional error of $0.002/0.122 \approx 2\%$. Any PN corrections that one might hope to resolve should contribute at least $2\%$ to the value of the ISCO. Using the test-mass limit as a gauge, the $n$th-order PN expansion of Eq.~\eqref{eq:Eschw} predicts fractional errors for the Schwarzschild ISCO of $\Omega^{n{\rm PN}}/6^{-3/2}-1 = [7.0,0.82,0.27,0.12,0.055,0.028]$ for $n=1$ to $6$. This indicates, for example, that the 5PN and higher-order terms account for $12\%$ of the ISCO, 6PN and higher-order terms account for $5.5\%$, etc. This suggests that the simulations in \cite{caudill-etal-initialdata-PRD2006} should be precise enough to resolve 4PN and 5PN effects, although not necessarily with high accuracy or without contamination from higher-PN terms.

We can attempt to also derive constraints on $e_4(0)$ and $e_5(0)$ using the BS conservative GSF corrections to the ISCO. In this case we compute the condition $d/dx [\tilde{E}^{\rm PN}_{\rm hybrid}{\bm (}\Omega=\Omega^{\rm isco, sf}_{\rm flat}(\eta){\bm )}]=0$, where we substitute Eq.~\eqref{eq:Omega_flat_renorm} for the frequency into a hybrid energy function analogous to Eq.~\eqref{eq:Ehybrid} [but with the 4PN and 5PN terms in Eq.~\eqref{eq:E3PN} included]. The resulting ISCO condition is then expanded to linear order in $\eta$, yielding the constraints
\bs
\begin{align}
\label{eq:e4cond2}
e_4(0) &\approx 429.1 \;\; \text{at 4PN order and} \\
e_4(0) &+ e_5(0)/5 \approx 382.8 \;\; \text{at 5PN order.}
\end{align}
\es

To further constrain the functions $e_4(\eta)$ and $e_5(\eta)$ we make the following two observations: first, if we examine the numerical values of the coefficients in the 3PN expansion of $\tilde{E}^{\rm PN}$ [cf.~Equation\eqref{eq:E3PN}],
\begin{multline}
\label{eq:E3pn-numerical}
\! -\frac{2\tilde{E}}{x} \approx 1 + x (-0.75 - 0.083 \eta) + x^2 (-3.4 + 2.4 \eta - 0.042 \eta^2) \\
+ x^3 (-11. + 39. \eta - 1.6 \eta^2 - 0.0068 \eta^3) + O(x^4),
\end{multline}
we see that the coefficients of each power of $\eta$ tend to decrease in absolute value as the power of $\eta$ increases. Second, since $\eta$ is at most $0.25$, the absolute value of the terms proportional to $\eta^p$ is further suppressed by a factor $0.25^p$ or smaller. This suggests that we can approximately ignore some of the higher-order $\eta$-terms in the expansions in Eqs.~\eqref{eq:e4e5-expand}. For example, working at the 4PN level only, we can approximate $e_4(\eta) \approx e_4^{(0)} + \eta e_4^{(1)}$. Then Eqs.~\eqref{eq:e4cond1} and \eqref{eq:e4cond2} imply
\be
\label{eq:e4-values}
e_4^{(0)} \approx 429.1 \;\;\;\; \text{and} \;\;\;\; e_4^{(1)} \approx -1127.
\ee
One could do something similar at the 5PN level, but as there are more unknown parameters than equations, one would need numerical values of the ISCO for more values of $\eta$. These values should be computable via QCID calculations analogous to those in \cite{caudill-etal-initialdata-PRD2006}, and could also help to constrain the higher-order coefficients $e_4^{(2)}$, $e_4^{(3)}$, etc. One might also suspect that the terms $e_4^{(0)}$ and $e_5^{(0)}$ could be constrained by current GSF calculations as was done with the redshift function $u_1^t(y)$ in \cite{blanchet-etal-selfforceII}.

The estimates on $e_4(\eta)$ and $e_5(\eta)$ presented here are meant to illustrate techniques through which they could be constrained. Better constraints would require more accurate numerical simulations for several mass ratios. Nonetheless, if the exact 4PN and 5PN terms are eventually computed, it would be interesting to compare their values with the above estimates.
\subsection{\label{sec:QCID}A suggested approach for numerically computing higher-order PN corrections to the gravitational-wave phasing}
The above subsections indicate that the recent GSF results for the conservative shift in the ISCO can better inform our knowledge of comparable-mass waveform templates. However, as they are currently limited to the first-order self-force, GSF calculations are constrained to only provide information about the $q\ll 1$ limit of these templates. Numerical relativity provides exact comparable-mass waveforms, but becomes severely limited by computational costs for $q\lesssim 1/10$. These costs are especially severe if one desires many cycles before the merger (in the regime where our current 3PN waveforms are starting to lose phase coherence). Here, I propose an alternative inspiral template-generation method based on calibrating higher-order PN terms with low-cost (or at least ``lower-cost'') NR simulations.

Match-filtering-based detection and analysis methods are most sensitive to the phase of the gravitational waveform. For quasicircular binaries, the phase of the Fourier transform of the GW signal $\Psi(f)$ is determined by two ingredients from PN theory, the orbital energy $E(x)$ and the energy flux [actually the luminosity ${\mathcal L}_{\rm gw}(x) = -\dot{E}$] as a function of the orbital frequency $M\Omega = x^{3/2}$,
\be
\label{eq:d2psidf2}
\frac{d^2 \Psi(f)}{df^2} = -2\pi \frac{dE/df}{{\mathcal L}_{\rm gw}(f)},
\ee
where $f=\Omega/\pi$ is the GW frequency. For nonspinning binaries, these two ingredients are currently known to 3.5PN order. Computing the 4PN and higher-order terms will be very difficult, and it is not clear if they will be computed before the first detections. While the 3.5PN order terms are sufficient for detecting GWs, knowledge of higher-order phase corrections will allow the recovery of more signal-to-noise at later times into inspiral, allowing expensive NR simulations to focus on the cycles very close to merger.

Rather than run standard NR evolutions to compute waveform templates for ``smallish'' ($q\sim 1/10$) mass ratios, I instead propose the following two-pronged strategy that involves using NR to calibrate undetermined parameters in standard (non-EOB) waveform templates. The first part involves computing higher-order corrections to the orbital energy $E(x)$. As shown above, the 4PN and 5PN logarithmic pieces of this function have been recently computed \cite{blanchet-etal-selfforceII}, but the nonlogarithmic terms are unknown. Using quasicircular initial data computations of the equal-mass ISCO and the GSF conservative ISCO shift, Sec.~\ref{sec:fitEpn} set some additional constraints on the undetermined functions at 4PN and 5PN orders. My primary suggestion is to use several QCID calculations---at a variety of mass ratios---to set further constraints on the undetermined 4PN and 5PN pieces of $E(x)$. While this could be accomplished by computing the ISCO for a variety of mass ratios, a better strategy might be to compute the energy for a variety of frequencies and mass ratios. This would essentially determine $E(x)$ up to 4PN or 5PN order through numerical fits to the QCID results. This is analogous to the fitting procedure used in \cite{blanchet-etal-selfforceII} to determine high-PN-order terms in the gauge-invariant redshift function $u_1^t$. If run for small enough mass ratios ($q\lesssim 1/100$), QCID calculations could potentially provide additional points of comparison with GSF calculations.

The second step is to ``compute'' the GW luminosity to higher PN order than 3.5PN. In this case we are partially helped by analytic BH perturbation theory calculations of the test-mass limit terms in the luminosity, which are currently known to 5.5PN order (see \cite{saski-tagoshi-LRR} and references therein). It might also be possible to extend the program of Ref.~\cite{blanchet-etal-selfforceII} to the computation of the finite-mass-ratio logarithmic terms in the luminosity at 4PN and higher orders. To obtain the remaining finite-mass ratio nonlogarithmic terms, one fits these undetermined coefficients by comparing with the luminosity from full NR evolutions. These evolutions are very expensive if the inspiral starts at large separations or has small mass ratios. However, to fit higher-PN terms in the luminosity we do not necessarily need full evolutions of the entire inspiral and merger. Rather, we could make do with small stretches of a simulation that consist of only a few orbits near a single frequency. This allows us to improve the fit to ${\mathcal L}_{\rm gw}(x,\eta)$ by supplementing the currently available NR values of ${\mathcal L}_{\rm gw}$ with a few discrete points at large separations and/or small mass ratios. Although even these few-orbit evolutions might still be expensive, they could have a big payoff in providing a permanent calibration of the PN phasing [determined via Eq.~\eqref{eq:d2psidf2}].

In practice, there are several difficulties associated with the above scheme. One obvious issue concerns the accuracy of current QCID calculations. Many (but not all) of these calculations assume that the spatial metric is conformally-flat (see \cite{bishop-etal-nonconformalID-PRD2004} and references therein); this is known to be accurate only to 1PN order. While some higher-order PN effects are still implicit in these calculations (which indeed show good agreement with 3PN calculations, cf.~Table \ref{tab:iscoequalmass}), one would clearly like a calculational scheme that is at least self-consistent to the order of the PN corrections that one is trying to compute (in this case 4PN). In computing the GW luminosity, there are also problems associated with performing NR simulations at large separations (and small mass ratios): evolutions are slow for these orbits, a large computational grid is required, and one must be careful of contamination from junk radiation and boundary reflections. One must also perform the calculations at separations large enough that the adiabatic approximation [upon which Eq.~\eqref{eq:d2psidf2} relies] is valid. However, at large separations the higher-order PN terms that one is trying to fit also become increasingly small, making them potentially difficult to resolve in a numerical simulation.

To gauge the needed precision, we can consider the size of the 4PN terms of $E(x)$ and ${\mathcal L}_{\rm gw}$ in the test-mass limit. The fractional error in the size of the 4PN term, ($E^{\rm 4PN}-E^{\rm 3PN})/E^{\rm Schw}$ varies from $\sim0.0002$ at $x=1/20$ to $\sim0.004$ at $x=1/10$ to $\sim0.03$ at $x=1/6$. For comparison, I estimate that the ISCO energy calculated in \cite{caudill-etal-initialdata-PRD2006} has a precision of roughly $1\%$ (see their Table II). This is just enough precision to resolve the 4PN term near the ISCO, but not enough to resolve it at much larger separations. Similarly, we can examine the PN expansion of the GW luminosity, which is known to 5.5PN order in the test-mass limit [see, e.g., Eq.~(174) of \cite{saski-tagoshi-LRR}]. In this case the fractional error $({\mathcal L}_{\rm gw}^{\rm 4PN}-{\mathcal L}_{\rm gw}^{\rm 3.5PN})/{\mathcal L}_{\rm gw}^{\rm 5.5PN}$ varies from $\sim0.001$ at $x=1/20$ to $\sim0.02$ at $x=1/10$ to $\sim0.1$ at $x=1/6$. Current NR simulations can attain this level of precision out to at least $x\sim1/10$ (see, e.g., Fig.~2 of \cite{boyle-etal-Efluxcomparison}). Future work will examine in more detail the feasibility of the scheme proposed here.
\section{\label{sec:conc}Conclusions}
The primary purpose of this study has been to compare the recent gravitational-self-force (GSF) calculations \cite{barack-sago_isco,barack-sago-circselfforcePRD2007,barack-sago-eccentricselfforce} of the conservative shift in the Schwarzschild ISCO to nearly all PN/EOB methods for computing the ISCO. The results, summarized in Table \ref{tab:compareisco}, show that while EOB methods calibrated to NR simulations perform best, uncalibrated EOB---as well as other resummation approaches---do not perform better than the gauge-invariant ISCO condition of \cite{blanchetiyer3PN}. This ISCO condition has the especially interesting property of exactly reproducing the Schwarzschild ISCO, even though it is derived using the standard PN equations of motion without any form of resummation (which is typically used to \emph{enforce} the test-mass limit). To investigate if this agreement is accidental, I have also generalized this gauge-invariant ISCO condition to spinning BHs, and showed that it reproduces the Kerr ISCO up to the expected order in the spin and PN expansion parameters. This approach also exactly reproduces the fully-relativistic conservative shift in the ISCO due to the spin of the test-mass (see \cite{favata-PNspinisco} for details).

The various PN/EOB ISCO methods were also compared with the quasicircular initial data calculations of the equal-mass ISCO in \cite{caudill-etal-initialdata-PRD2006}. In this case, a nonresummed PN method also performs better than EOB approaches. However, this is a different method (one based on the minimum of the 3PN orbital energy) than the Blanchet-Iyer \cite{blanchetiyer3PN} ISCO condition. The 3PN-EOB approach has the advantage of being a \emph{single} resummed method that can model the ISCO in both the comparable-mass and test-mass limits with comparable (albeit larger) errors in both cases.

These results suggest that the standard PN equations of motion somehow contain information about the strong-field conservative dynamics (at least in the test-mass limit). This is surprising since PN quantities often converge slowly in the $\eta\rightarrow 0$ limit. The gauge-invariant ISCO condition of \cite{blanchetiyer3PN} (and its generalization to spinning binaries) apparently does not suffer from this limitation.
Subsequent work will examine predictions of the conservative ISCO shift in Kerr from EOB and PN approaches \cite{favata-PNspinisco}. These predictions can be compared with future GSF calculations in the Kerr spacetime.

The $\approx 28\%$ error between the 3PN EOB prediction for the conservative GSF ISCO shift and the exact Barack-Sago result suggests that while the EOB formalism exactly encapsulates the test-particle limit of motion in Schwarzschild, it does not encapsulate small-deviations from the test-mass limit any better than standard PN approaches. While this is not necessarily unexpected, it suggests caution be used when attempting to model conservative GSF effects in EMRI waveforms using EOB methods \cite{yunes-etal-EMRI-EOB}. Of course, the EOB formalism can be modified via the introduction of unknown terms that can be calibrated to GSF calculations \cite{damour-GSF,barack-damour-sago_periastron} or to numerical relativity. However, for intermediate mass ratios there is no accurate numerical method to calibrate against, so it is useful to compare the performance of different approaches in the absence of any calibration.

This study has also explored how GSF calculations can further our knowledge of comparable-mass template waveforms. GSF calculations of the conservative ISCO shift can be used to calibrate parameters in EOB and phenomenological IMR waveforms, and, combined with quasicircular initial data calculations of the comparable-mass ISCO, can help constrain the undetermined functions in the 4PN and 5PN pieces of the PN orbital energy. Calculations of quasicircular initial data sequences at unequal mass ratios are needed to further constrain these parameters and functions.

A new method of calibrating 4PN (or higher) terms in the waveform phasing was also proposed. Rather than comparing with the full-NR waveform phase, this approach suggests using two separate NR calculations to determine the phase at a specific frequency: quasicircular initial data sequences can be used to determine the orbital energy at specific orbital frequencies, while full-NR evolutions at large separations (but for a small number of orbits) can determine the energy flux at (nearly) specific frequencies. These two calculations are presumably less costly than a long numerical evolution, and they provide the necessary ingredients to determine the GW phase as a function of frequency. This scheme will be further explored in future work.
\begin{acknowledgments}
This research was supported through an appointment to the NASA Postdoctoral Program at the Jet Propulsion Laboratory, administered by Oak Ridge Associated Universities through a contract with NASA. I gratefully acknowledge Emanuele Berti, Luc Blanchet, Alessandra Buonanno, and Alexandre Le Tiec  for detailed comments on this manuscript. For helpful discussions I thank Parameswaran Ajith, Curt Cutler, Harald Pfeiffer, Mark Scheel, Michele Vallisneri, Bernard Whiting, and participants of the ``Theory meets data analysis at comparable and extreme mass ratios'' conference (Perimeter Institute, June 2010). I also thank the anonymous referee for helpful comments that improved this manuscript.
\end{acknowledgments}
\bibliography{text_favata_isco-studyV2b}
\end{document}